\documentclass[12pt]{article}

\usepackage{epsfig,multicol,multirow}
\usepackage{amsmath,subfigure,latexsym,amssymb,cite}

\newcommand{\be}{\begin{equation}}
\newcommand{\ee}{\end{equation}}
\newcommand{\nn}{\nonumber}
\newcommand{\bea}{\begin{eqnarray}}
\newcommand{\eea}{\end{eqnarray}} 

\newcommand{\la}{\langle}
\newcommand{\ra}{\rangle}

\newcommand{\Z}{\mathbb{Z}}
\newcommand{\R}{{\kern+.25em\sf{R}\kern-.78em\sf{I} \kern+.78em\kern-.25em}}
\newcommand{\RR}{{\kern+.25em\sf{R}\kern-.6em\sf{I} \kern+.6em\kern-.25em}}
\newcommand{\N}{{\kern+.25em\sf{N}\kern-.78em\sf{I} \kern+.78em\kern-.25em}}
\newcommand{\C}{{\kern+.25em\sf{C}\kern-.50em\sf{I} \kern+.50em\kern-.25em}}

\newcommand{\ri}{{\rm i}}

\newcommand{\vp}{\varphi}

\makeatletter
\@addtoreset{equation}{section}
\makeatother

\begin{document}
 
\begin{center}

{\Large\bf Measuring the Topological Susceptibility}

\vspace*{6mm}

{\Large\bf in a Fixed Sector} \\

\vspace*{1cm}

Irais Bautista$^{\rm a,b}$, Wolfgang Bietenholz$^{\rm a}$, 
Arthur Dromard$^{\rm c}$, \vspace*{1mm} \\ 
Urs Gerber$^{\rm a}$,  Lukas Gonglach$^{\rm c}$, Christoph P.\ Hofmann$^{\rm d}$, 
\vspace*{1mm} \\
H\'{e}ctor Mej\'{\i}a-D\'{\i}az$^{\rm a}$ and Marc Wagner$^{\rm c}$ \\
\ \\

{\small
$^{\rm \, a}$  Instituto de Ciencias Nucleares \\
Universidad Nacional Aut\'{o}noma de M\'{e}xico \\
A.P. 70-543, C.P. 04510 Distrito Federal, Mexico \\
\ \\ \vspace{-2mm}

$^{\rm \, b}$ Facultad de Ciencias F\'{\i}sico Matem\'{a}ticas, Benem\'{e}rita 
Universidad \\
Aut\'{o}noma de Puebla, A.P. 1364, Puebla, Mexico\\
\ \\ \vspace{-2mm}

$^{\rm \, c}$ Goethe-Universit\"{a}t Frankfurt am Main\\
Institut f\"{u}r Theoretische Physik \\
Max-von-Laue-Stra\ss e 1, D-60438 Frankfurt am Main, Germany\\
\ \\ \vspace{-2mm}

$^{\rm \, d}$ Facultad de Ciencias, Universidad de Colima\\
Bernal D\'{\i}az del Castillo 340, C.P. 28045 Colima, Mexico
}
\end{center}

\vspace*{6mm}

\noindent
For field theories with a topological charge $Q$, it is often of 
interest to measure the topological susceptibility $\chi_{\rm t}
 = (\la Q^{2} \ra -\la Q \ra^{2}) / V$.
If we manage to perform a Monte Carlo simulation where $Q$ changes
frequently, $\chi_{\rm t}$ can be evaluated directly.
However, for local update algorithms and fine
lattices, the auto-correlation time with respect to $Q$ tends to
be extremely long, which invalidates the direct approach.
Nevertheless, the measurement of $\chi_{\rm t}$ is still feasible,
even when the entire Markov chain is topologically frozen.
We test a method for this purpose, based on the correlation of the
topological charge density, as suggested by Aoki, Fukaya, Hashimoto
and Onogi. Our studies in non-linear $\sigma$-models and in 2d Abelian
gauge theory yield accurate results for $\chi_{\rm t}$, which confirm 
that the method is applicable. 
We also obtain promising results in 4d SU(2) Yang-Mills theory,
which suggest the applicability of this method in QCD.

\section{Motivation}

We are going to address the functional integral formulation of quantum
physics in Euclidean space with periodic boundary conditions. 
For a number of models of interest, the
configurations occur in distinct topological sectors, each one characterized 
by a topological charge $Q \in \Z$. Examples are 2d Abelian 
gauge theory and 4d Yang-Mills theories.
In these cases also fermions may be present, so this class includes
the Schwinger model and QCD. Further examples are the O($N$)
models (non-linear $\sigma$-models) in $N-1$ dimensions, and all
2d CP($N-1$) models.

We deal with the case where parity symmetry holds, which implies 
the expectation value $\la Q \ra =0$. Then the topological
susceptibility is given by
\be  \label{chitdef}
\chi_{\rm t} = \int d^{d}x \, \la q(0) q(x) \ra =
\frac{\la Q^{2} \ra}{V_{\rm cont}} \ ,
\ee
where $q$ is the topological charge density ($Q =  \int d^{d}x \, q(x)$),
and $V_{\rm cont}$ is the volume. This quantity is often of interest; 
for instance $\chi_{\rm t}$ of quenched QCD is relevant
for the Witten-Veneziano relation \cite{WV}.
Clearly, $\chi_{\rm t}$ can only be determined on the 
non-perturbative level. Hence lattice simulations are the 
appropriate method for this purpose. (Actually the lattice definitions
of $q$ and $Q$ are slightly ambiguous, see {\it e.g.}\ Refs.\ \cite{Marc}
for comparative studies; we will specify later the formulations
that we use.)

Here we consider the 1d O(2) model (quantum rotor), 
the 2d O(3) model (Heisenberg model), as well as 2d U(1),
and 4d SU(2) gauge theories.
In our Monte Carlo study of non-linear $\sigma$-models, we
apply a cluster algorithm \cite{Wolff}, which performs non-local 
update steps. Hence it frequently changes the topological sector,
so it provides precise results for $\chi_{\rm t}$ by direct measurements.

In most other models of quantum field theory, especially in
almost all models with fermions or gauge fields, such an
efficient algorithm is not known. There one resorts 
to local update algorithms, such as the heatbath algorithm,
which we used in our gauge theory simulations.
In that case, the Markov chain tends to get stuck in one 
topological sector, in particular as one approaches the 
continuum limit. That may well happen in lattice QCD
with light dynamical quarks and a lattice spacing
below $0.05~{\rm fm}$ \cite{openbc}.

In light of these prospects for the near future, indirect methods 
to measure $\chi_{\rm t}$ are of interest. Here we test systematically
the Aoki-Fukaya-Hashimoto-Onogi (AFHO) method \cite{AFHO},
which evaluates $\chi_{\rm t}$ based on the density correlations $\la 
q_{0} \, q_{x}\ra_{|Q|}$, measured at fixed $|Q|$.
Hence this quantity enables the determination of $\chi_{\rm t}$ 
even from a Markov chain that is entirely confined to a single
topological sector. 

An alternative concept with the same motivation is sketched in
Ref.\ \cite{Philippe}. For a recent study with a related  
approach, see Ref.\ \cite{MaxLike}. 
The procedure of Ref.\ \cite{BCNW} is more general, but it
contains yet another option to determine $\chi_{\rm t}$ from
topologically restricted measurements.

Here we are going to demonstrate in a variety of models that the 
AFHO method works. For suitable settings, it provided $\chi_{\rm t}$ 
values which are correct to 2 or 3 digits. 
We will also discuss the practical limitations of this approach.

\section{Topological charge density correlation}

The AFHO method was derived
in Ref.\ \cite{AFHO}, inspired by related considerations
by Brower, Chandrasekharan, Negele and Wiese \cite{BCNW}.
It deals with the long-distance correlation 
of the topological charge density $q_{x}$ at fixed $|Q|$. 
The topological susceptibility $\chi_{\rm t}$
can be evaluated from the (approximate) relation
\bea
^{~ \lim}_{x \to \infty} \ \langle q_{0} \, q_{x} \rangle_{|Q|} & \approx &
-\frac{1}{V^{2}} \left( \la Q^{2} \ra - Q^{2} + 
\frac{V c_{4}}{2 \la Q^{2} \ra} \right) \nn \\
& = & - \frac{\chi_{\rm t}}{V} + 
\frac{1}{V^{2}} \left( Q^{2} - \frac{c_{4}}{2 \chi_{\rm t} } \right) \ ,
\label{denseq}
\eea
where $V$ is the volume in lattice units. The term
\be  \label{kurto}
c_{4} = \frac{1}{V} \Big( 3 \langle Q^{2} \rangle^{2} -
\langle Q^{4} \rangle \Big)
\ee
is the kurtosis, which measures the deviation from a Gaussian 
distribution of the topological charges. It tends to be tiny, 
see {\it e.g.}\ Refs.\ \cite{topGaus} for quenched QCD results,
and in the 1d O(2) model it vanishes exactly in the continuum 
and infinite volume \cite{rot97}. In the current context
its contribution can be ignored, as we will see in the following.

Eq.\ (\ref{denseq}) consists of the leading terms of an expansion
in $1/\la Q^{2} \ra$, therefore $\la Q^{2} \ra = V \chi_{\rm t}$ 
should be large. Since $\chi_{\rm t}$ is expected to stabilize 
in the large volume limit, eq.\ (\ref{denseq}) holds up to 
sub-leading finite size effects. Moreover,
its derivation assumes the ratio $|Q| / \la Q^{2} \ra$ to be small,
hence it is favorable to apply this method in sectors of small $|Q|$.

With these assumptions, eq.\ (\ref{denseq}) shows that the correlation 
of the topological charge density in a fixed sector is not expected 
to vanish over long distances.
Instead it is expected to attain a plateau, which 
depends on $|Q|$: it is slightly negative
for $Q=0$ (obviously, a fluctuation of $q_{0}$ has to be
compensated elsewhere), but it rises for increasing $Q^{2}$.

The AFHO method has been tested previously in the 2-flavor Schwinger 
model with light chiral fermions \cite{BHSV}. The numerically 
measured correlations $\la q_{0} \, q_{x} \ra_{0}$ suggest that a 
conclusive evaluation of $\chi_{\rm t}$ requires large statistics:
on a $16 \times 16$ lattice it requires $O(10^{5})$ configurations. 
Variants of this method were already applied in 2-flavor QCD, though with a 
different density \cite{chitQCDNf2}, and recently also in QCD 
with $2+1$ flavors, with a reduction to sub-volumes \cite{JLQCD14}.
For a precise test, the non-linear $\sigma$-models are perfectly suited, 
since the method can be probed with high statistics, and the results
for $\chi_{\rm t}$ can be compared with reliable direct measurements. 
In order to probe the potential of this approach further, we add 
investigations in 2d Abelian and 4d non-Abelian gauge theories.
Synopses of this study were anticipated in proceeding contributions 
\cite{procs}.

\section{Results for the 1d O(2) model}

We start with the 1d O(2) model, or 1d XY model,
which describes a quantum mechanical 
particle moving freely on the circle $S^{1}$, with periodic boundary 
conditions in Euclidean time $x$.
In continuous time, a trajectory can be described by an angle $\varphi (x)$, 
with $\varphi (0) = \varphi (L_{\rm cont})$. On the lattice we deal with 
angles $\varphi_{x}$, $x= 1 \dots L$ and 
$\varphi_{L+1} = \varphi_{1}$.\footnote{All lattice quantities will be
given in lattice units.}
We introduce the nearest site difference
\be  \label{deltaphi}
\Delta \vp_{x} = (\vp_{x+1} - \vp_{x}) \ {\rm mod} \ 2 \pi \in
( - \pi, \pi ] \ ,
\ee
{\it i.e.}\ the modulo function is defined such that $|\Delta \vp_{x}|$
is minimized.

This is one of the simplest models with a topological charge, 
which is given by
\be \label{QQ}
Q = \frac{1}{2\pi} \int_{0}^{L_{\rm cont}} dx \, \vp '(x) \ \ {\rm (continuum)} 
\ , \quad
Q = \frac{1}{2\pi} \sum_{x=1}^{L} \Delta \vp_{x} \ \ {\rm (lattice)} \ , \ \ 
\ee
where $q(x) = \vp '(x)/(2 \pi)$ is the topological charge density in the
continuum, and  $q_{x} = \Delta \vp_{x} /(2 \pi)$ is its geometrically
defined counterpart on the lattice.

The continuum action reads
$S_{\rm cont}[\vp ] = \frac{\beta_{\rm cont}}{2} 
\int_{0}^{L_{\rm cont}} dx\, \vp ' (x)^{2}$. 
In Appendix A we show that relation (\ref{denseq}) without 
the kurtosis term, 
\be  \label{qq1dO2}
\la q(0) q(x) \ra_{|Q|} = - \frac{\chi_{\rm t}}{V} + 
\frac{Q^{2}}{V^{2}} \ ,
\ee
is exact in this case, and independent of the separation $x$.

In our numerical study, we consider three lattice actions: 
the standard action, the Manton action \cite{Manton} and the 
constraint action
\cite{constraint},
\bea
S_{\rm standard}[\vp ] &=& \beta \sum_{x=1}^{L} (1 - \cos \Delta \vp_{x}) 
\ , \quad
S_{\rm Manton}[\vp ] = \frac{\beta}{2} \sum_{x=1}^{L} \Delta \vp_{x}^{2} 
\ , \quad \nn \\
S_{\rm constraint}[\vp ] &=& \left\{ \begin{array}{ccc}
0 && \Delta \vp_{x} < \delta \quad \forall x \\
+ \infty && \mbox{otherwise} \end{array} \right. \ .
\label{latact1dO2}
\eea
The parameter $\beta$ (or $\beta_{\rm cont}$) corresponds here to the 
moment of inertia, and in the last case $\delta$ is the constraint angle.

\begin{table}[h!]
\begin{center}
\begin{tabular}{|l||c|c|}
\hline
action & $\xi$ & $\chi_{\rm t}$ \\
\hline
\hline
 && \vspace*{-4mm} \\
continuum  & $2 \beta_{\rm cont}$ & $\frac{1}{4 \pi^{2} \beta_{\rm cont}}$ \\
 && \vspace*{-4mm} \\
\hline
 && \vspace*{-4mm} \\
standard & $ \left[ \ln \frac{\int_{-\pi}^{\pi} d \vp \ 
\exp (- \beta (1-\cos \vp))}{\int_{-\pi}^{\pi} d \vp \ 
\exp (- \beta (1-\cos \vp)) \cos \vp} \right]^{-1} $
      & $\frac{1}{4 \pi^{2}} 
\frac{\int_{-\pi}^{\pi} d \vp \ \vp^{2} \exp (- \beta (1 - \cos \vp ))}
{\int_{-\pi}^{\pi} d \vp \ \exp (- \beta (1 - \cos \vp ))}$ \\
 && \vspace*{-4mm} \\
\hline
 && \vspace*{-4mm} \\
Manton     & $ \left[ \ln \frac{\int_{-\pi}^{\pi} d \vp \ 
\exp (- \beta \vp^{2} / 2 )}{\int_{-\pi}^{\pi} d \vp \ 
\exp (- \beta \vp^{2}/2 ) \cos \vp} \right]^{-1} $ & 
$\frac{1}{4 \pi^{2}} \frac{\int_{-\pi}^{\pi} d \vp \ \vp^{2} 
\exp (- \beta \vp^{2} / 2 )} 
{\int_{-\pi}^{\pi} d \vp \ \exp (- \beta \vp^{2} / 2 ))}$\\
\hline
 && \vspace*{-4mm} \\
constraint & $ \left[ \ln (\delta / \sin (\delta))\right]^{-1} $ 
& $\frac{\delta^{2}}{12 \pi^{2}} $ \\
\hline
\end{tabular}
\end{center}
\vspace*{-3mm}
\caption{Closed expressions for the correlation length $\xi$ and
the topological susceptibility $\chi_{\rm t}$ in the 1d O(2) model
at infinite size, in the continuum and for three lattice actions.}
\label{xichit}
\end{table}

\begin{figure}[h!]
\begin{center}
\includegraphics[width=0.35\textwidth,angle=270]{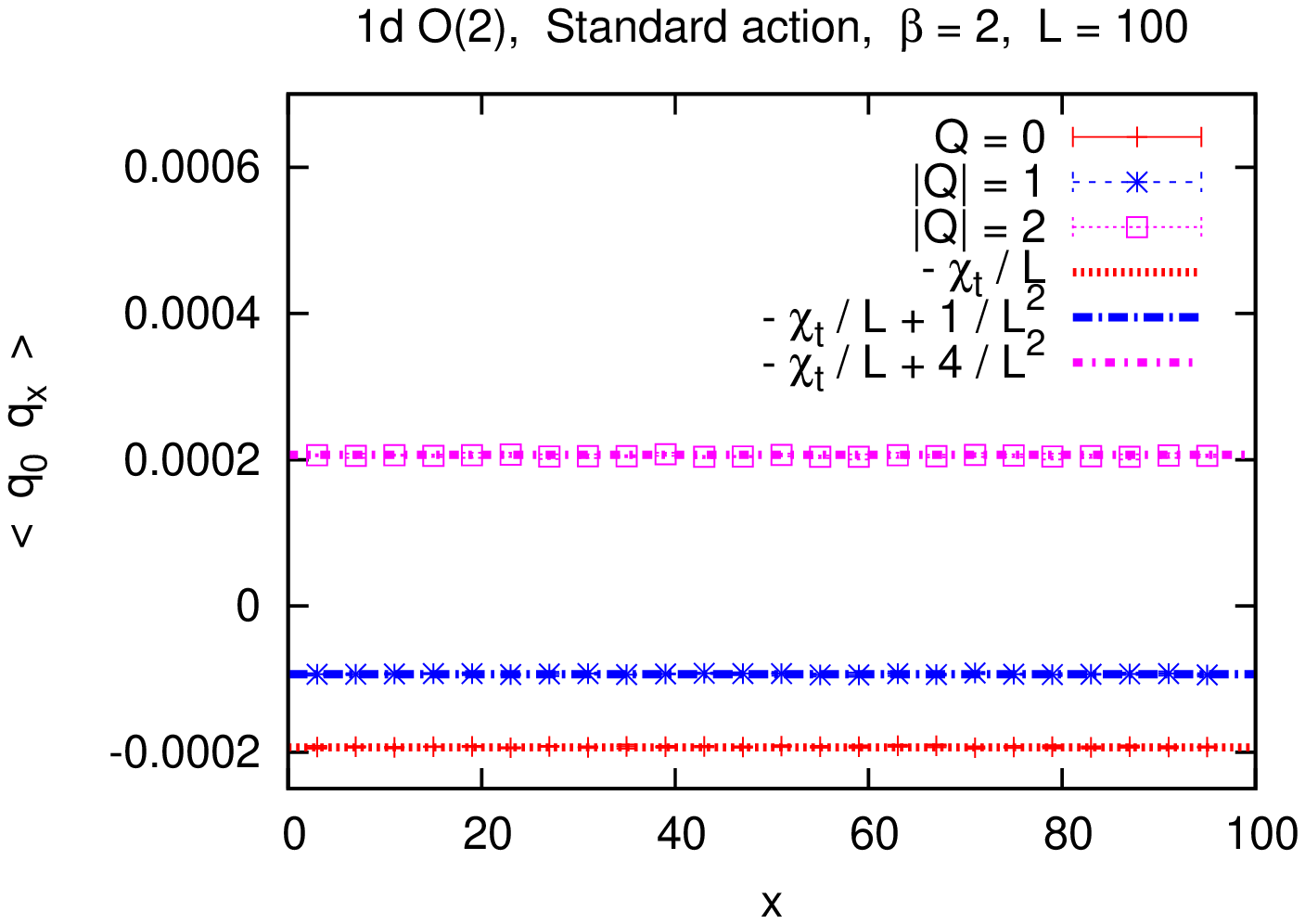}
\includegraphics[width=0.35\textwidth,angle=270]{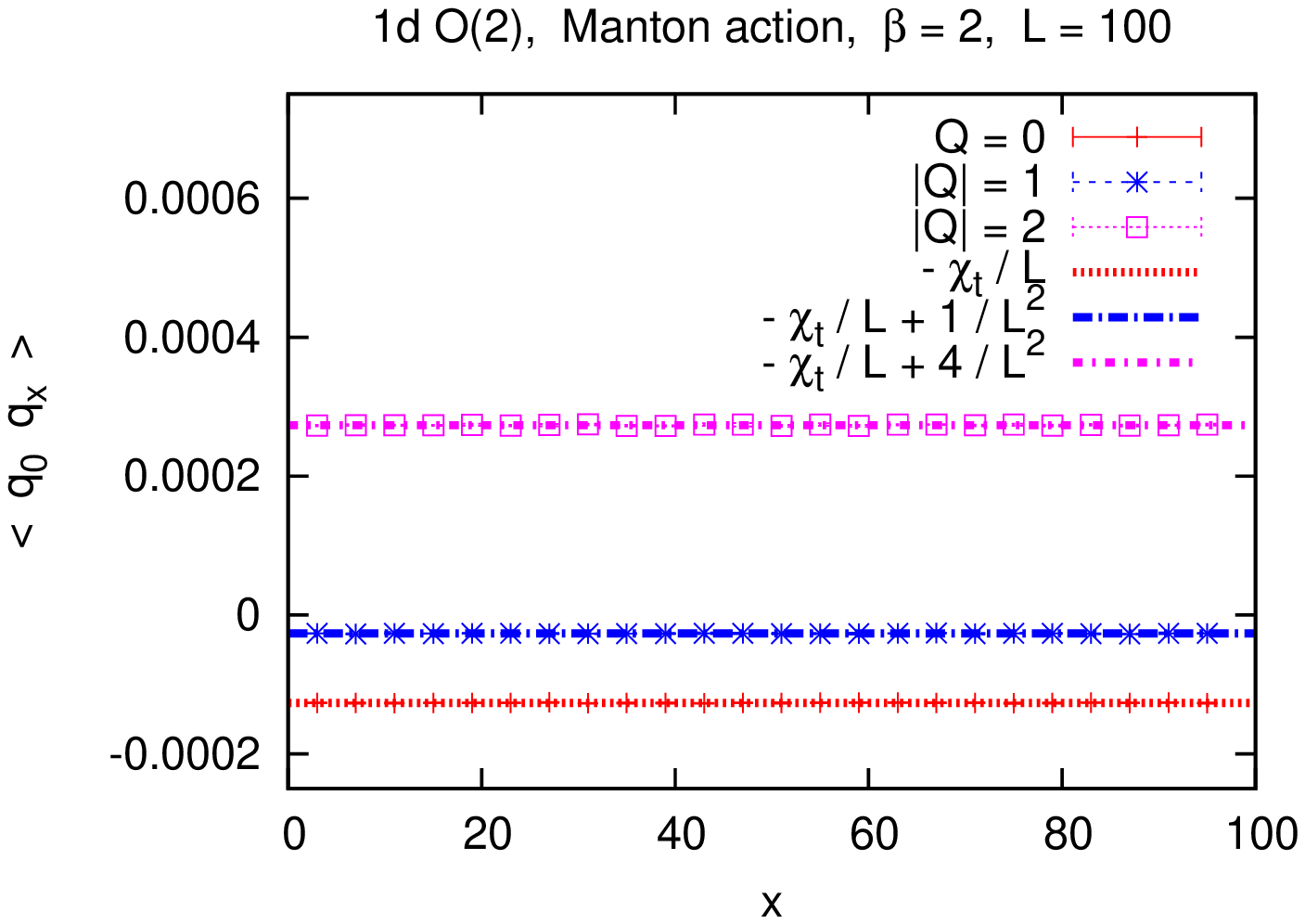}
\includegraphics[width=0.35\textwidth,angle=270]{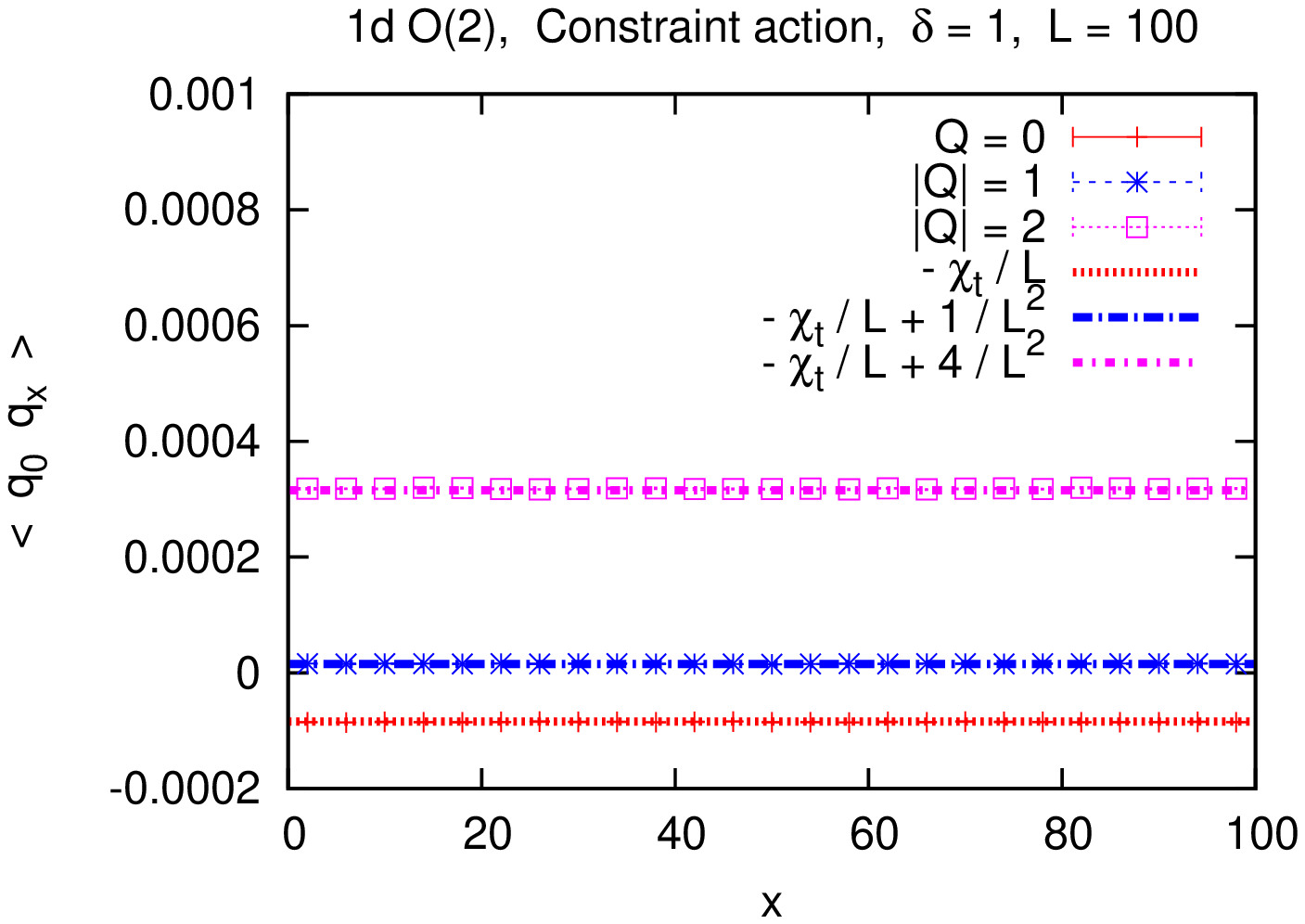} 
\caption{The topological charge density correlation in the 1d O(2) 
model over a distance of $x$ lattice spacings, at $L =100$. The first
two plots refer to the standard action and the Manton action
at $\beta =2$, with $\xi = 2.779$, $\la Q^{2} \ra = 1.936$,
and $\xi = 4.000$, $\la Q^{2} \ra = 1.266$, respectively.
The plot below is obtained with the constraint action at 
$\delta =1$, $L=100$, $\xi = 5.793$, $\la Q^{2} \ra = 0.844$.
For comparison, we include in all cases horizontal lines for the 
prediction based on eq.\ (\ref{qq1dO2}), where we insert the directly 
measured values of $\chi_{\rm t}\,$.}
\vspace*{-5mm}
\label{1dO2SMCqq}
\end{center}
\end{figure}

In the thermodynamic limit, $L \to \infty$, the correlation 
length $\xi = 1 / (E_{1} - E_{0})$ ({\it i.e.}\ the inverse energy gap) 
and $\chi_{\rm t}$ are known analytically \cite{rot97,constraint}, 
as we summarize in Table \ref{xichit}.\footnote{For 
the continuum action, Ref.\ \cite{Arthur} discusses the spin correlation 
function, and its restriction to a single topological sector.}
The product $\xi \, \chi_{\rm t}$ is a scaling quantity, {\it i.e.}\
a dimensionless term composed of observables.
In the continuum it amounts to 
\be
\xi \, \chi_{\rm t}|_{\rm continuum} = \frac{1}{2 \pi^{2}} \ .
\ee
This value is attained for the lattice actions in the limit
$\beta \to \infty$ and $\delta \to 0$, respectively, which reveals 
a facet of universality even in one dimension. The corresponding scaling 
behavior is discussed in Refs.\ \cite{rot97,rot07,constraint}. 
In particular, the Manton action scales excellently, since it 
is classically perfect.

Figure \ref{1dO2SMCqq} shows examples for 
numerically measured correlations $\la q_{0} \, q_{x} \ra_{|Q|}$,
using these actions at $L=100$. We see in all cases
that the numerical data are in excellent
agreement with the predicted plateau values. These plateaux
are accurately visible, so the AFHO method does indeed enable a 
precise numerical determination of $\chi_{\rm t}$.

\begin{figure}[h!]
\begin{center}
\includegraphics[width=0.4\textwidth,angle=270]{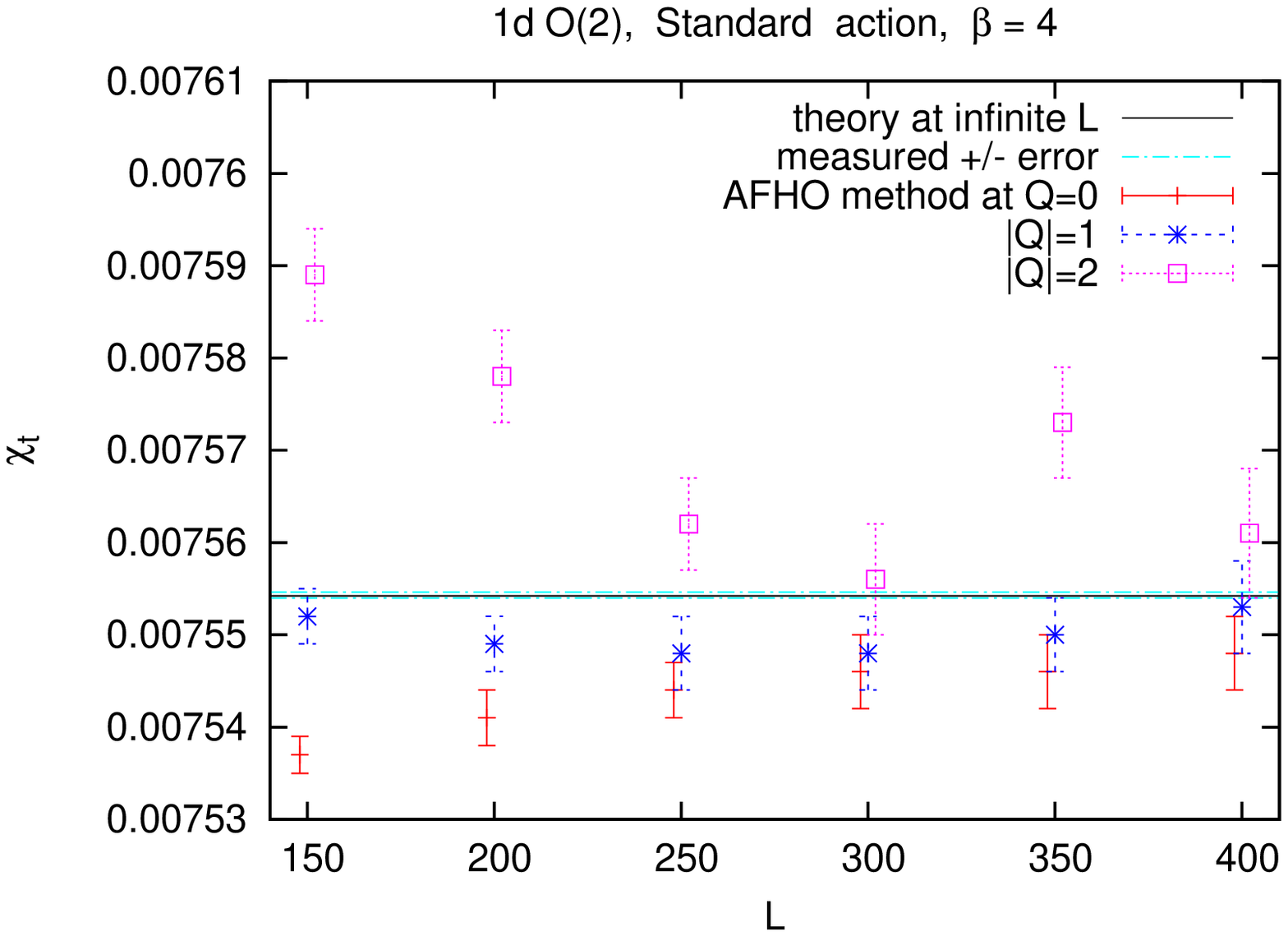}\\
\includegraphics[width=0.4\textwidth,angle=270]{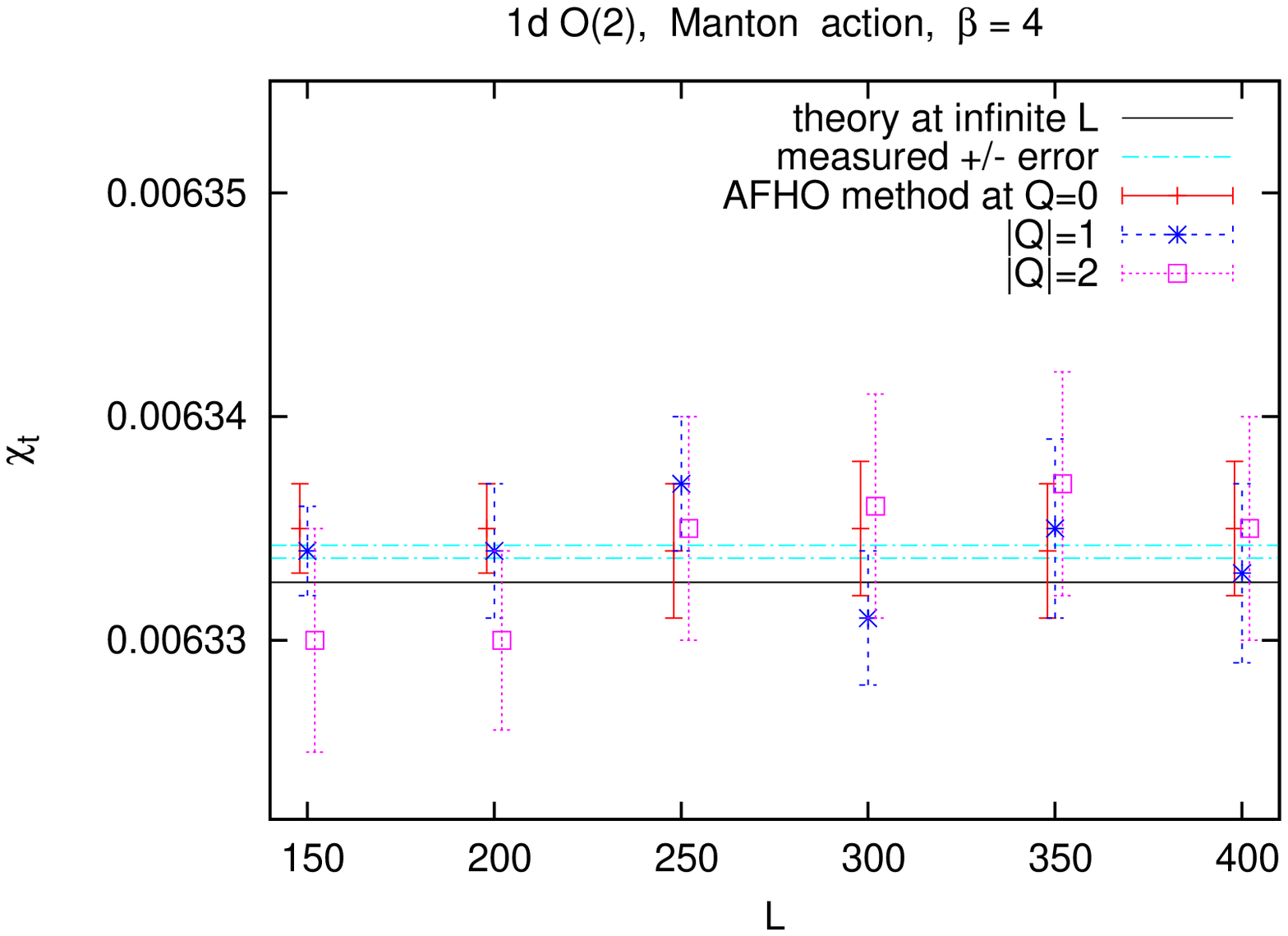}
\caption{The topological susceptibility $\chi_{\rm t}$ in 
the 1d O(2) model for the standard and Manton action at $\beta=4$,
where $\xi =6.8$ and $8.0$, respectively.
We show the theoretical value at
$L = \infty$, the directly measured value (at $L=400$), and the
values obtained from the AFHO method in the range 
$L=150 \dots 400$, in the sectors $|Q| = 0,\ 1,\ 2$.
For the standard action, there are permille level 
deviations from the predicted value, in particular for $|Q|=2$, which 
are suppressed for increasing $L$. For the Manton action all results 
coincide to an impressive precision, even down to $L=150$.}
\vspace*{-7mm}
\label{1dO2B4qq}
\end{center}
\end{figure}

To demonstrate this explicitly, we consider
the range $L= 150 \dots 400$, and $|Q| =0,\ 1,\ 2$, 
which leads to the results for $\chi_{\rm t}$
in Figure \ref{1dO2B4qq}. For the Manton action we obtain precise 
agreement with the theoretical $\chi_{\rm t}$ value in all cases.
For the standard action we observe small deviations up to a few 
permille, which are suppressed for $|Q| \leq 1$, and for $|Q|=2$
they are reduced as $L$ increases.

These tiny lattice artifacts are revealed due to
extremely large statistics: for each parameter set, at least $ 5 \cdot
10^{9}$ measurements have been performed with a cluster algorithm. 
This yields very precise results, and illustrates the convergence 
towards the theoretical $\chi_{\rm t}$ value for increasing 
$\la Q^{2}\ra$ (or equivalently $L$).

\section{Results for the 2d O(3) model}

We proceed to field theory, and first to the 2d O(3)
model (or Heisenberg model), with periodic boundary conditions. 
In its lattice formulation a classical spin
variable of unit length is attached to each lattice site $x$,
$\vec e_{x} \in S^{2}$. 

Regarding the topological charge, we consider sets of three 
neighboring spins. In our case the lattice consists of quadratic
plaquettes, and each plaquette is divided into two triangles
(the cutting diagonal has an alternating orientation between nearest 
neighbor plaquettes).
Each of these triangles carries such a set of spins
$\vec e_{1}, \, \vec e_{2}, \, \vec e_{3}$.
They are connected on the sphere $S^{2}$ by the
arcs of minimal length to form a spherical triangle.
Its oriented area $A$ is given by \cite{constraint}
\bea
x &=& 1 + \vec e_{1} \cdot \vec e_{2} + \vec e_{2} \cdot \vec e_{3}
+ \vec e_{1} \cdot \vec e_{3} \ , \quad y = \vec e_{1} \cdot 
(\vec e_{2} \times \vec e_{3}) \ , \nn \\ 
\varphi &=& {\rm arg} (x + {\rm i} y) \ \mbox{mod} \ 2 \pi \ , \quad
A (\vec e_{1},\vec e_{2}, \vec e_{3}) = 2 \varphi \ ,
\eea
where the modulo function is defined as in eq.\ (\ref{deltaphi}).
In each plaquette, the normalized sum of these two oriented
spherical triangles (the total solid angle),
$q_{x} = A ( \vec e_{x},\vec e_{x + \hat 1},
\vec e_{x + \hat 2}, \vec e_{x + \hat 1 + \hat 2} ) / (4 \pi)$,
is the topological charge density.

If we sum over all plaquettes, and thus  over all triangles, 
we obtain the geometrically defined topological charge
\be
Q = \sum_{x} q_{x} =
\frac{1}{4 \pi} \sum_{x} A ( \vec e_{x},\vec e_{x + \hat 1},
\vec e_{x + \hat 2}, \vec e_{x + \hat 1 + \hat 2} )
\in \Z \ .
\ee
This definition, which was advocated in Ref.\ \cite{BergLuscher}, has 
the virtue of providing integer $Q$ values for all configurations (except
for a subset of measure zero), just like eq.\ (\ref{QQ}). It counts how
many times (and with which orientation) these triangles cover the sphere. 

The standard lattice action reads
\be
S_{\rm standard} [ \vec e \, ] = \beta \sum_{x, \mu} (1 - \vec e_{x} \cdot
\vec e_{x + \hat \mu} ) \ ,
\ee
where $\mu$ runs from 1 to 2, $\hat \mu$ is the unit vector
in $\mu$-direction, and $\beta > 0$.

\begin{figure}[h!]
\begin{center}
\hspace*{-6mm}
\includegraphics[width=0.41\textwidth,angle=270]{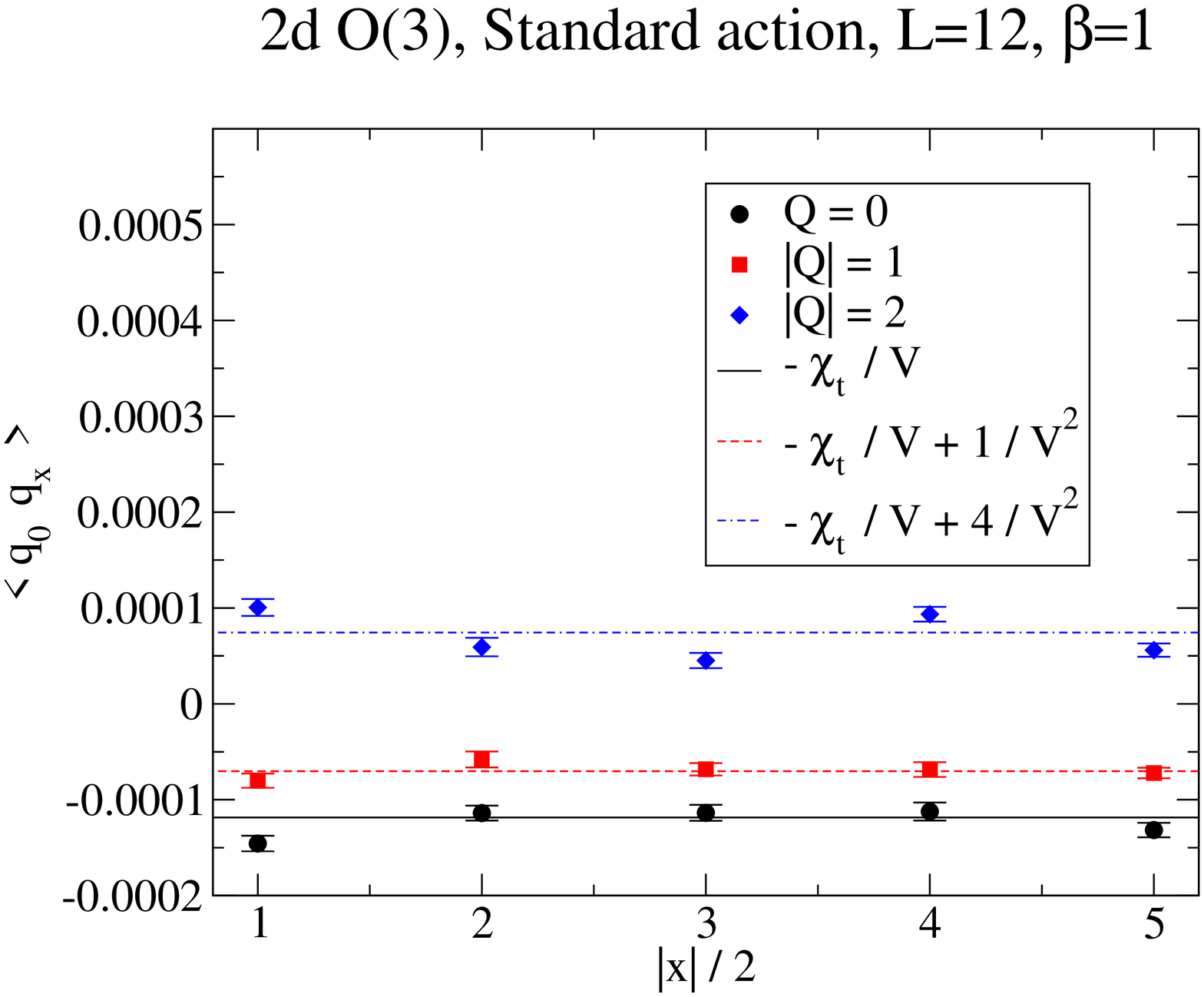}
\hspace*{-8mm}
\includegraphics[width=0.41\textwidth,angle=270]{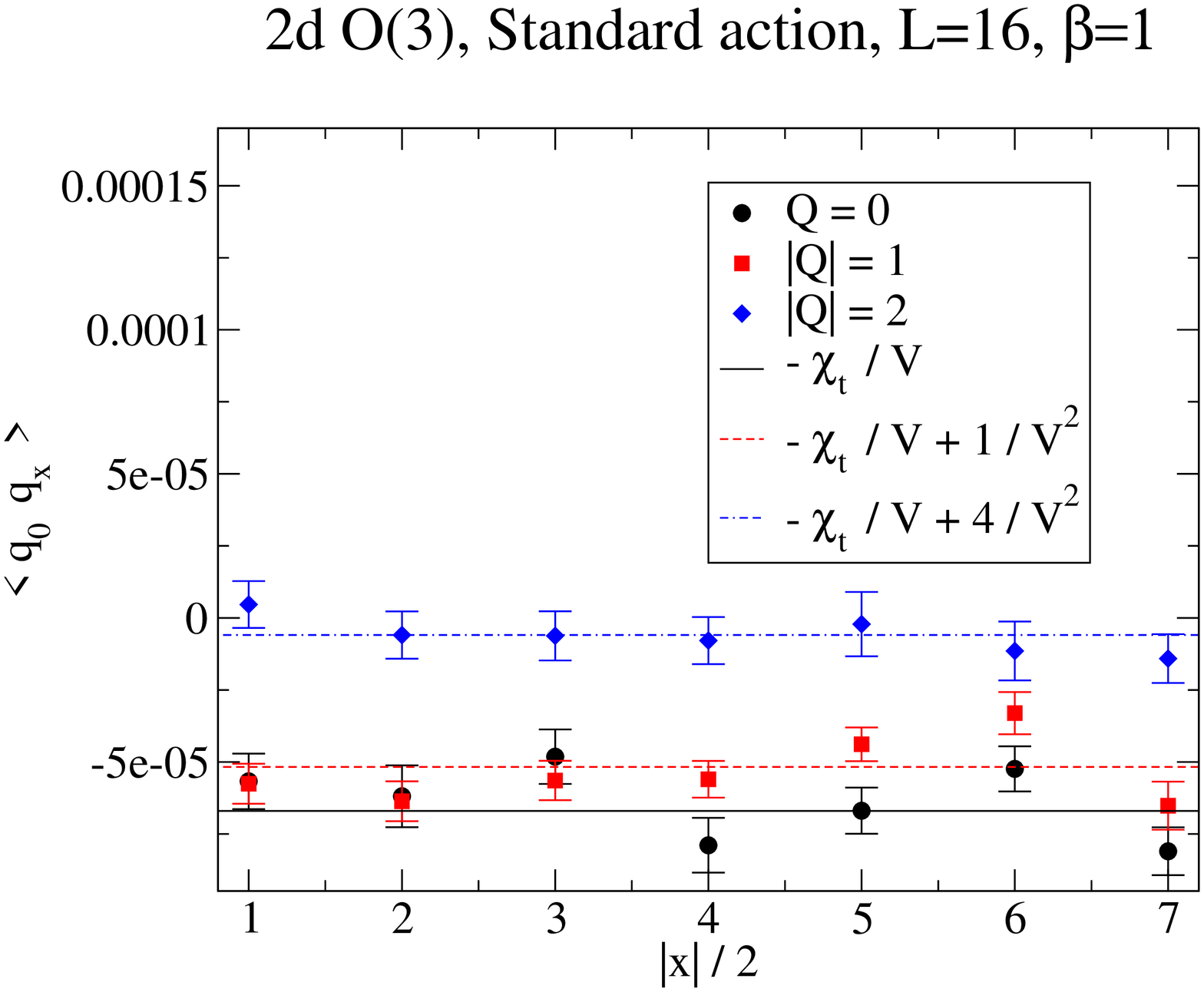} 
\vspace*{-4mm}
\caption{The topological charge density correlation in the 2d O(3)
model, with the standard lattice action at $\beta = 1$.
We show data measured on $L \times L$
lattices, $L=12$ and $16$, in the sectors $|Q|= 0, \, 1, \, 2$. They
are compared to lines for the values according to eq.\ (\ref{qq1dO2}), 
with the directly measure susceptibility $\chi_{\rm t}$.}
\vspace*{-3mm}
\label{fig2dO3stanB1}
\end{center}
\end{figure}

Figure \ref{fig2dO3stanB1} shows the topological charge density 
correlation $\la q_{0} \, q_{x}\ra$ at $\beta =1$, measured in the 
sectors $|Q| = 0, \ 1,\ 2$, on $L \times L$ lattices of size $L = 12$ 
and $L = 16$. The measurements are carried out parallel to the axes, and
the spin separation proceeds in steps of two lattice units, due to
the alternating triangularization in the definition of $q_{x}$.
The horizontal lines are the expected plateau values according
to eq.\ (\ref{qq1dO2}). Again we inserted the directly measured 
values of $\chi_{\rm t} = \la Q^{2} \ra /V$;
they are very precise, thanks to the use of a cluster algorithm,
which provided a statistics of $O(10^{7})$ well 
thermalized and decorrelated measurements.\footnote{This model 
is sometimes considered topologically ill, 
because $\chi_{\rm t} \, \xi^{2}$, which is supposed
to be the scaling term, diverges logarithmically in the continuum 
limit. In the integral representation of eq.\ (\ref{chitdef}),
this effect emerges at distance $x=0$; at finite distances,
the topological charge density correlation is a controlled 
quantity \cite{constraint}. Here we determine $\chi_{\rm t}$ 
with different methods at fixed $\xi$, so this
defect does not affect our study.} 

The plots clearly confirm the qualitatively expected picture.
We further confirm that the condition
of a large separation is quite harmless: 
for the separation of four lattice spacings, the plateau value 
is already well attained. 

\begin{figure}[ht!]
\begin{center}
\includegraphics[width=0.45\textwidth,angle=270]{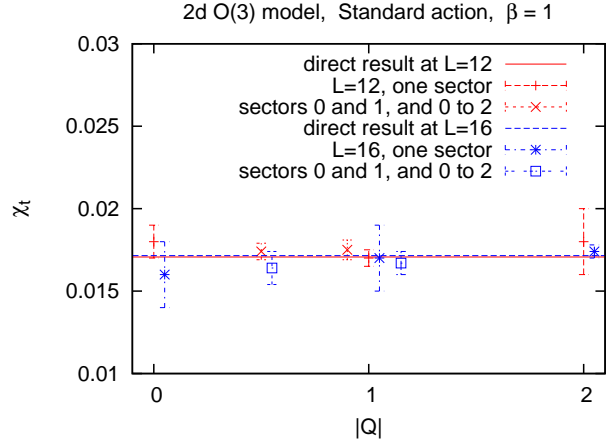}
\end{center}
\vspace*{-3mm}
\caption{The topological susceptibility in the 2d O(3) 
model on $12 \times 12$ and $16 \times 16$ lattices, 
with the standard action at $\beta=1$. 
We show results obtained from fits to the topological charge 
density correlation in sector $|Q|$, or combined fits in several 
sectors. This is compared to the direct measurement
as $\la Q^{2} \ra /V$, which is very precise.
All results are compatible within the errors.}
\label{tab2dO3stanB1}
\end{figure}

For a quantitative analysis, we perform individual fits of the data 
to a constant in one sector (skipping $2 \dots 6$ points at the 
boundaries); each fit yields a value of the topological 
susceptibility $\chi_{\rm t}$. 
In addition we consider combined fits in two or three sectors.
These results are confronted with the directly measured values
in Figure \ref{tab2dO3stanB1}. 
As theory predicts, the lowest $|Q|$-sectors are most reliable.
In fact, the evaluation of $\chi_{\rm t}$ based on 
$\la q_{0} \, q_{x}\ra_{|Q|}$ is 
successful to an accuracy of a few percent in these cases. However,
Figure \ref{fig2dO3stanB1} also shows
that the application of this method is getting difficult
when $L$ increases: then the values of $\la q_{0} \, q_{x}\ra_{|Q|}$
become tiny, and thus hard to distinguish from zero, and from each other.

\begin{figure}[h!]
\begin{center}
\hspace*{-6mm}
\includegraphics[width=0.41\textwidth,angle=270]{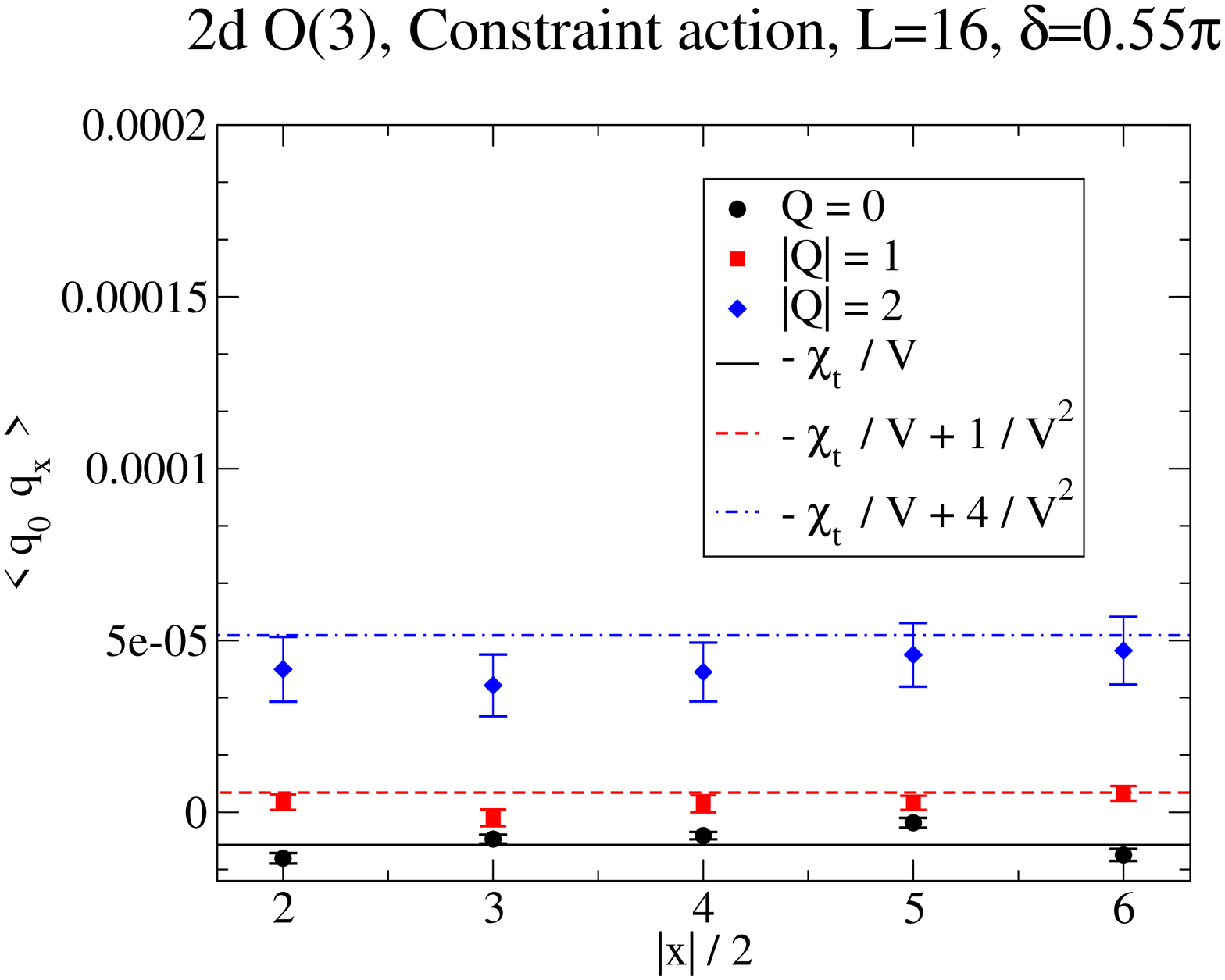}
\hspace*{-8mm}
\includegraphics[width=0.41\textwidth,angle=270]{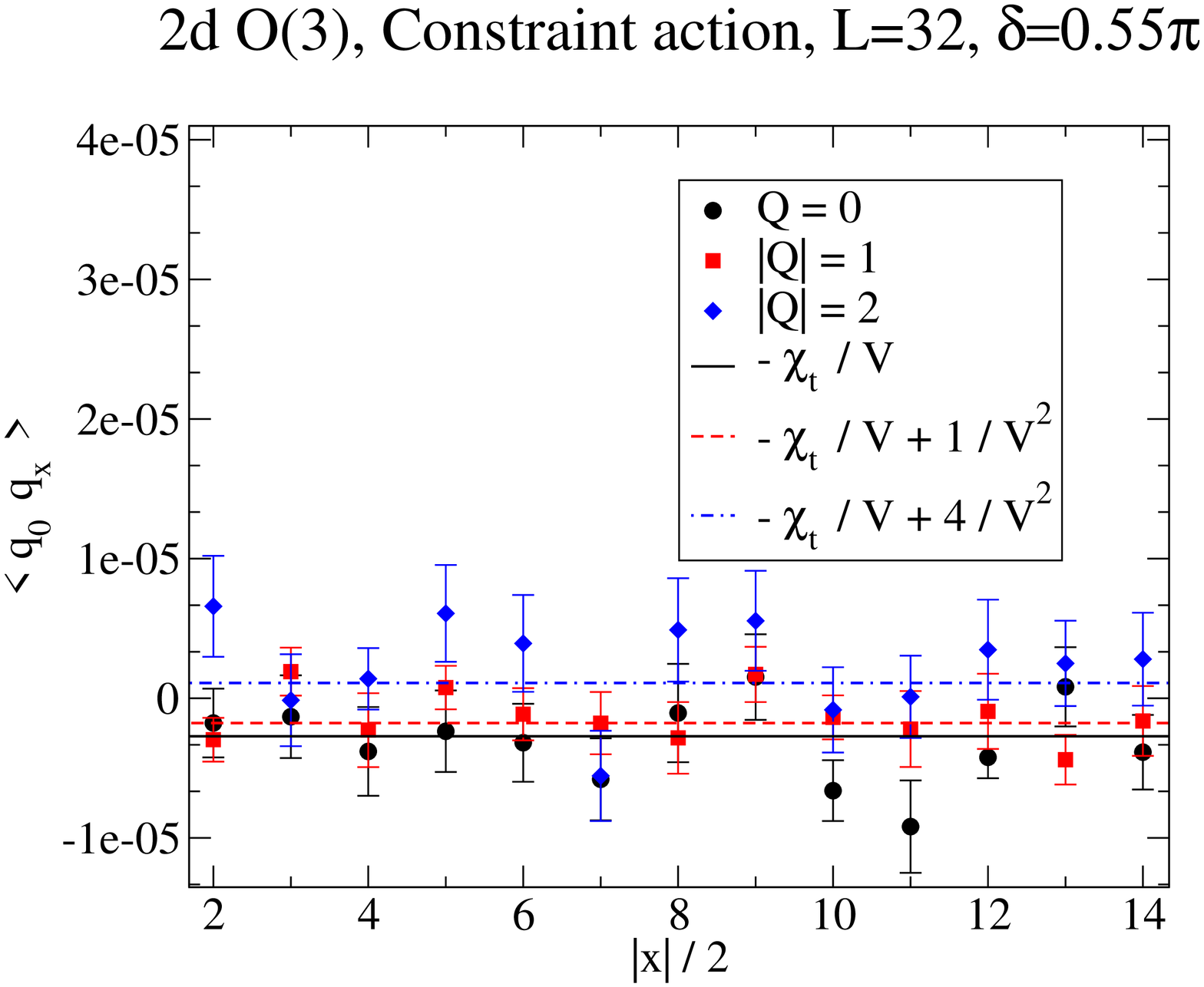} 
\vspace*{-4mm}
\caption{Plots analogous to Figure \ref{fig2dO3stanB1}, now
for the constraint action at $\delta = 0.55 \, \pi$, which corresponds
to $\xi \simeq 3.6$, and with lattice sizes $L=16$ and $L=32$.}
\vspace*{-3mm}
\label{fig2dO3constraint}
\end{center}
\end{figure}

\begin{figure}[ht!]
\begin{center}
\includegraphics[width=0.45\textwidth,angle=270]{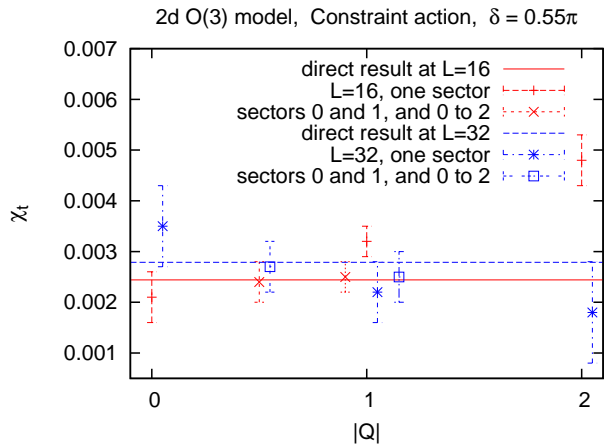}
\end{center}
\vspace*{-3mm}
\caption{The topological susceptibility in the 2d O(3) model 
on $16 \times 16$ and $32 \times 32$ lattices, with the constraint 
action at $\delta = 0.55 \, \pi$, where $\xi \simeq 3.6$. 
We display results from fits to the 
topological charge density correlation in sector $|Q|$, or combined 
fits in several sectors. Comparison to the direct measurement shows 
that the AFHO method is less successful than in the examples of
Figure \ref{tab2dO3stanB1}, since larger volumes are involved.}
\label{tab2dO3c}
\end{figure}

This is a case of strong coupling; $\beta =1$ leads to
a correlation length of $\xi \simeq 1.3$ (at large $L$).
Hence the volumes that we used can be considered large, but the lattice 
is coarse. In order to probe the AFHO method closer to the continuum
limit, we now proceed to a different setting. We move to the
constraint action, which is defined in analogy to eq.\ 
(\ref{latact1dO2}), {\it i.e.}\ the action is zero if all angles 
between neighboring spins are less than $\delta$, 
and infinite otherwise \cite{constraint}.
We set the constraint angle to $\delta = 0.55 \, \pi$,
which corresponds to a correlation length of $\xi \simeq 3.6$.
Accordingly, we now consider larger square lattices, with $L=16$ and $L=32$.
Figure \ref{fig2dO3constraint} shows the topological charge correlations
in this case, and Figure \ref{tab2dO3c} displays the fit results.

These plots show again that the condition of a larger separation 
$|x|$ is not a practical problem. However, despite the statistics 
of $O(10^{7})$ measurements, the AFHO method does run into 
trouble in reproducing the directly measured 
$\chi_{\rm t}$ values beyond one digit. This is mostly a consequence
of the larger volumes involved; they suppress the signal, which
is relevant to extract $\chi_{\rm t}$ in this indirect manner.
 
\begin{figure}[h!]
\begin{center}
\hspace*{-6mm}
\includegraphics[width=0.55\textwidth,angle=270]{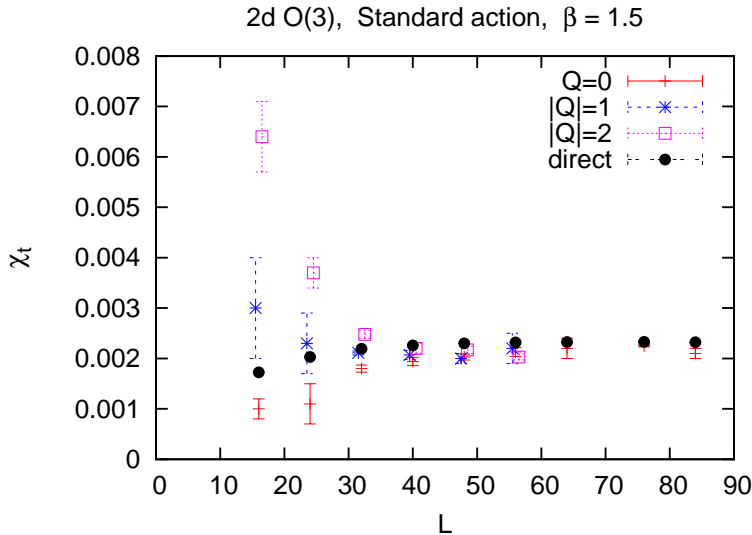}
\vspace*{-2mm}
\caption{The topological susceptibility $\chi_{\rm t}$ for the standard 
action at $\beta =1.5$, on $L \times L$ lattices with $L=16 \dots 84$
(with $\xi \simeq 9.5$ at large $L$).
The directly measured values stabilize for $L \geq 32$
(its errors are too small to be visible in this plot), and
the AFHO results approximate it well in the regime $L= 40 \dots 84$.
For smaller $L$, this method suffers from significant finite size effects,
and for larger $L$ the signal for the determination of $\chi_{\rm t}$ is 
too small for a good numerical resolution.}
\vspace*{-3mm}
\label{fig2dO3standardB15}
\end{center}
\end{figure}

Let us finally take a large step to a numerical experiment
very close to the continuum limit: it is performed with the
standard lattice action at $\beta =1.5$, on square lattices 
with $L= 16 \dots 128$; at large $L$, this corresponds
to $\xi \simeq 9.5$.
Figure \ref{fig2dO3standardB15} shows the results for $\chi_{\rm t}$ 
up to $L=84$, based on the topological charge correlations
in the sectors $|Q| = 0, \, 1,\, 2$, and by direct measurement.
As $L$ increases, the latter converges well at $L \geq 32$. 
The results by the AFHO method move towards the directly 
measured value, and get close to it at $L=40$. Here the 
range $L \approx 40 \dots 60$ is optimal for its application.
As we increase $L$ further, we face again the
problem that the tiny signal, which matters for $\chi_{\rm t}$,
gets lost in the statistical noise. For $|Q| = 1,\, 2$ this
happens already at $L \geq 64$; only at $Q=0$ the method
still leads to useful results up to $L=84$. (For completeness 
we add that at $L=128$ we obtained $\chi_{\rm t} = 0.0019(8)$, 
which is still compatible with the directly measured value 
$0.002292(9)$, but it has an error of $42 ~\%$).

\section{Results for 2d Abelian gauge theory}

We proceed to 2d U(1) gauge theory with the plaquette action,
{\it i.e.}\ Wilson's standard lattice formulation \cite{CreutzMM}.
We simulate it with the heatbath algorithm on periodic $L\times L$
lattices, with a variety of $L$ and $\beta$ values, which keep
$\la Q^{2} \ra$ in the range of $0.7$ to $10.4$. In each case,
the statistics involves $10^{7}$ configurations.
For the topological charge density $q_{x}$ we also applied the
straight plaquette regularization of the field strength
tensor in terms of non-compact link variables $A_{x,\mu}$,
\be
q_{x} = F_{x,12} = \frac{1}{2\pi} \Big[ (A_{x,1} + A_{x + \hat 1,2}
- A_{x + \hat 2,1} - A_{x,2}) \ {\rm mod} \ 2 \pi \Big] \ ,
\ee
still with the modulo function as defined in eq.\ (\ref{deltaphi}).
As in Section 4, its correlation was 
measured parallel to one of the axes.

Thanks to a generous separation of the measurements, 
the direct evaluation of $\chi_{\rm t}$ is very precise,
although the updates are local. 

\begin{figure}[h!]
\begin{center}
\includegraphics[width=0.36\textwidth,angle=270]{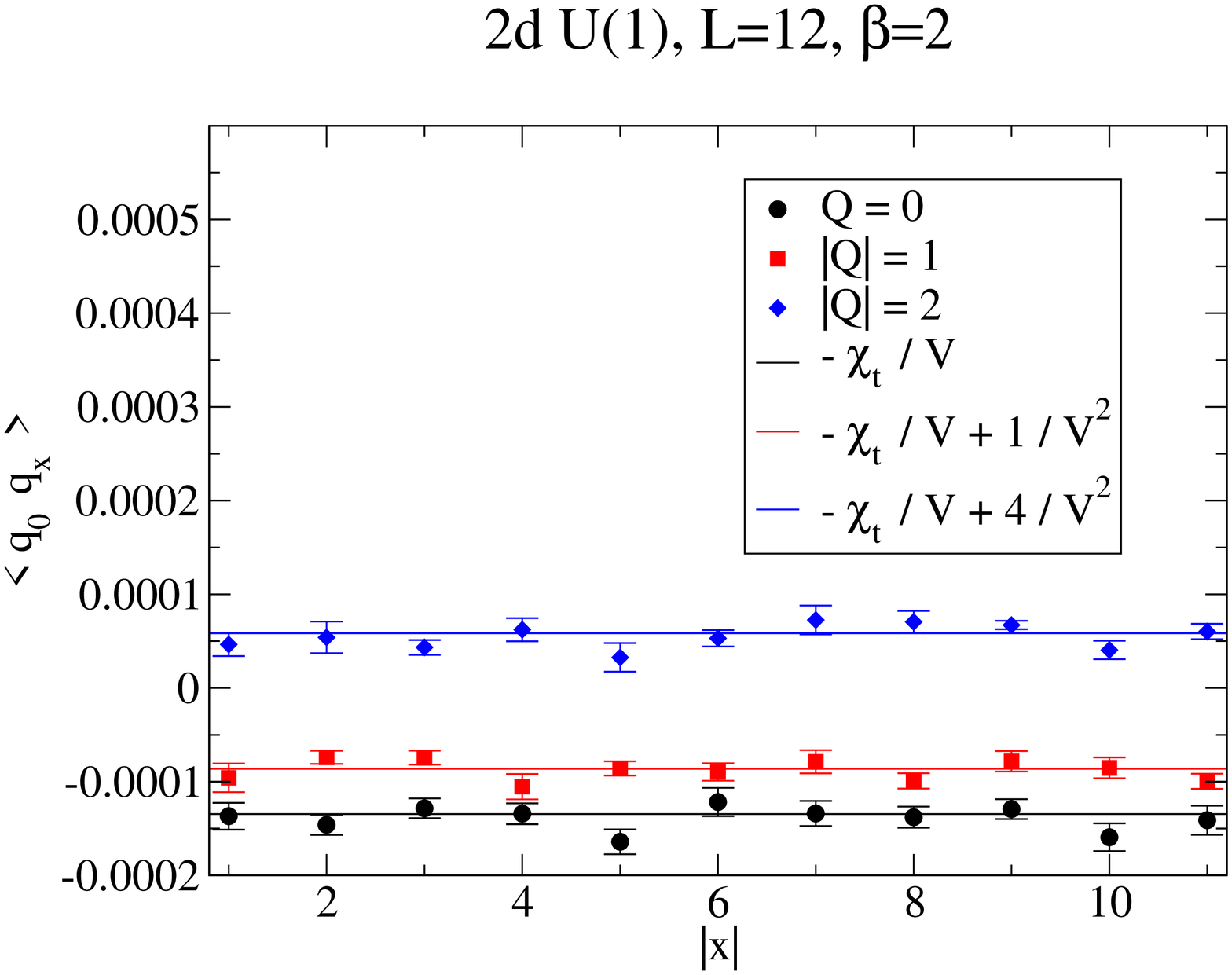}\\
\includegraphics[width=0.36\textwidth,angle=270]{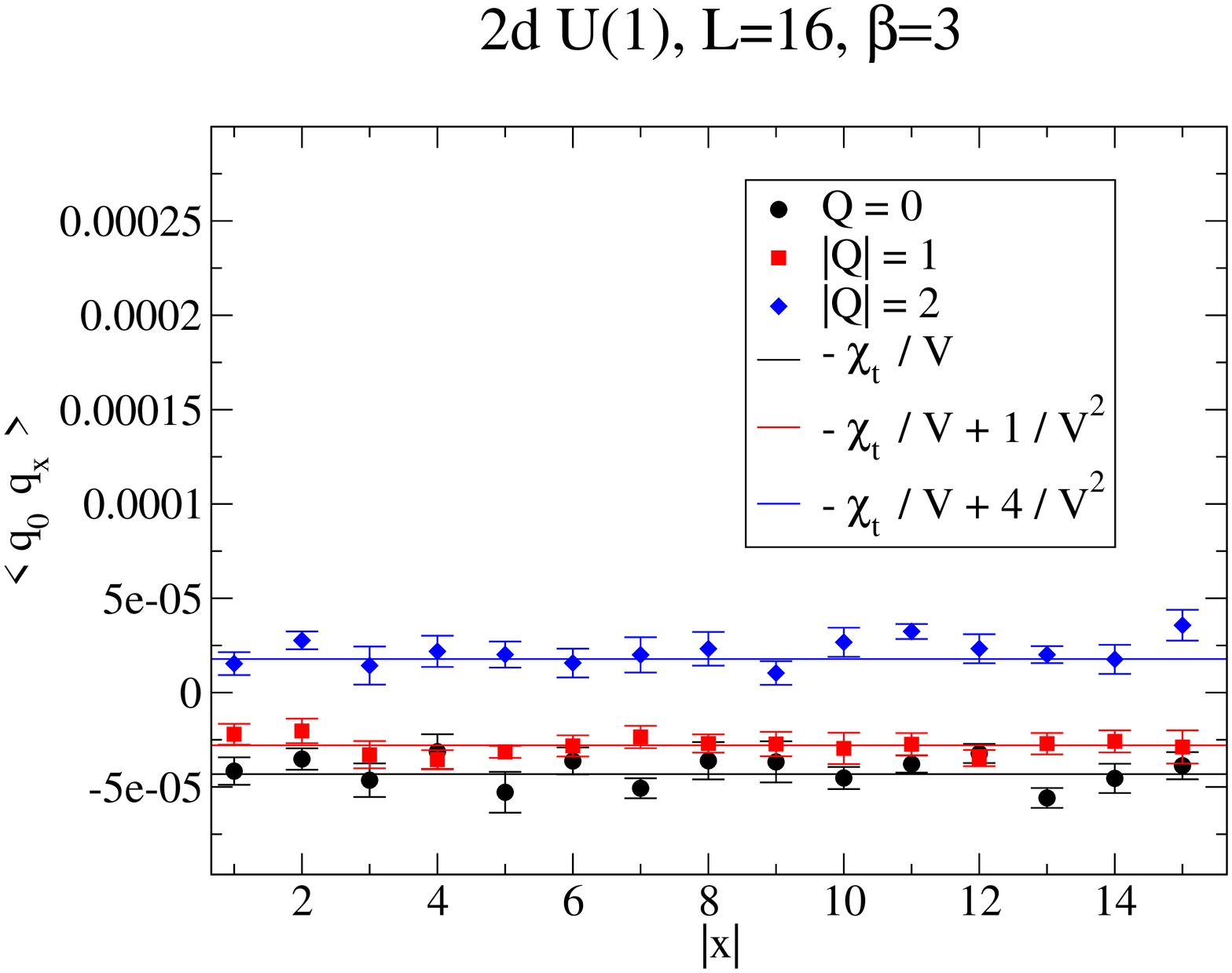}\\
\includegraphics[width=0.36\textwidth,angle=270]{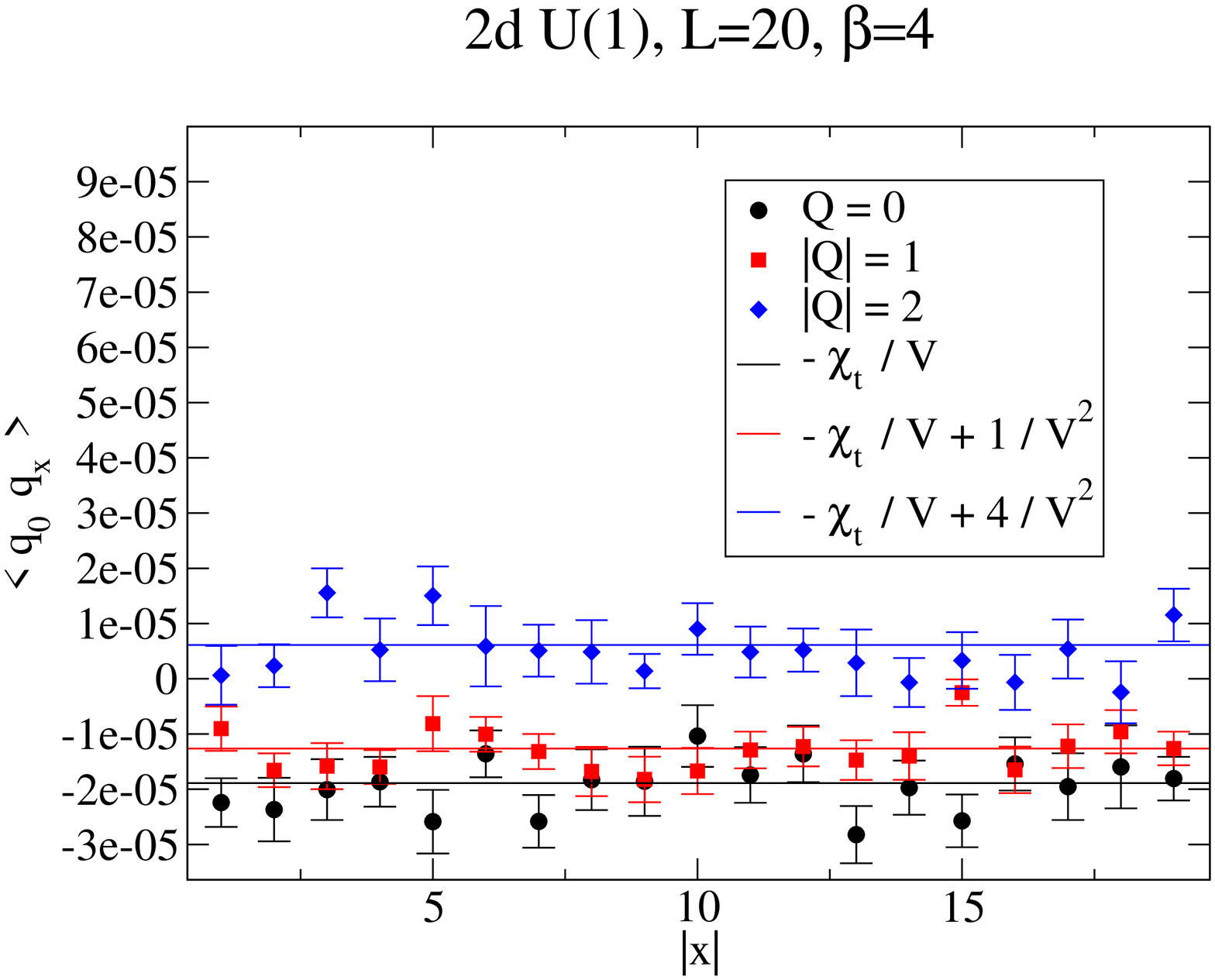} 
\caption{The topological charge density correlation in 2d U(1) 
gauge theory on $L \times L$ lattices, over a distance of $x$ 
lattice spacings. We show the first three plateau values (horizontal lines, 
based on direct measurements of $\chi_{\rm t}$), and the corresponding 
data for $\beta =2$, $L=12$, with a plaquette value of $\la P \ra = 0.6978$
and $\la Q^{2} \ra = 2.79$; 
$\beta =3$, $L=16$, $\la P \ra = 0.8100$, $\la Q^{2} \ra = 2.83$, and 
$\beta =4$, $L=20$, $\la P \ra = 0.8635$,  $\la Q^{2} \ra = 3.02$.}
\vspace*{-5mm}
\label{2dU1plat}
\end{center}
\end{figure}

Figure \ref{2dU1plat} gives three examples which illustrate that also 
here the data for the topological charge density match the predicted 
plateaux well, so that the determination of $\chi_{\rm t}$ by the
AFHO method is possible. In addition we see again the difficulty
setting in as the volume increases.

\begin{figure}[h!]
\begin{center}
\includegraphics[width=0.55\textwidth,angle=270]{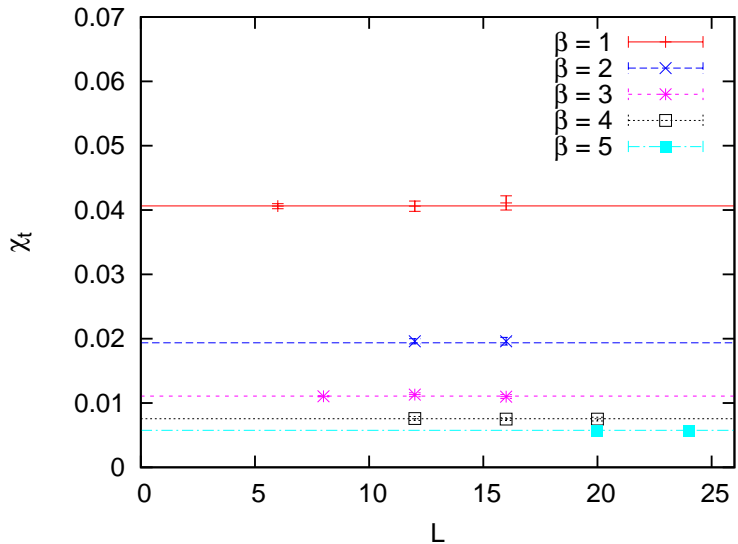}
\vspace*{-2mm}
\caption{The topological susceptibility $\chi_{\rm t}$ for 2d U(1)
gauge theory, on $L \times L$ lattices. The horizontal lines are 
the directly measured values (errors and differences for various
$L$ are invisible on this scale). The symbols are the AFHO values
obtained by a combined fit in the sectors $|Q| \leq 1$.}
\vspace*{-3mm}
\label{2dU1all}
\end{center}
\end{figure}

Figure \ref{2dU1all} provides an overview over the results
in the range $L = 6 \dots 20$ and $\beta = 1 \dots 5$. In the extreme
cases, the plaquette values amount to $\la P \ra = 0.4464$ ($\beta =1$) 
and $0.8934$ ($\beta =5$); for the rest we refer to the caption of
Figure \ref{2dU1plat}.

In this case, we apply the AFHO method as a combined fit to the
data in the topological sectors with $Q=0$ and $|Q|=1$. This
might be the optimal application, which includes a large
number of data points, but avoids the less reliable topological
sectors. In all the cases shown in Figure \ref{2dU1all}, these AFHO 
results for $\chi_{\rm t}$ agree with the directly measured values, 
within errors on the percent level, {\it e.g.}\ 
at $L=16$ we obtain $\chi_{\rm t} = 0.0196(6) \ (\beta =2)$; \
$0.0110(3) \ (\beta =3)$; \ $0.0075(2) \ (\beta =4)$.

\begin{figure}[h!]
\begin{center}
\includegraphics[width=0.5\textwidth,angle=270]{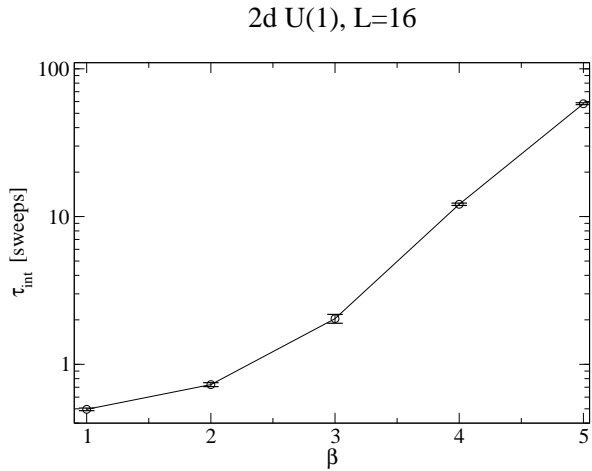}
\vspace*{-2mm}
\caption{The integrated autocorrelation time $\tau_{\rm int}$ in 
2d U(1) gauge theory, with respect to the topological charge $Q$. 
We consider  $\beta = 1 \dots 5$ on a $16 \times 16$ lattice, 
and measure $\tau_{\rm int}$ based on $10^{7}$ sweeps, after 
thermalization. As $\beta$ is getting large, we observe an
exponential increase of $\tau_{\rm int}$ (measured in the number
of sweeps).}
\vspace*{-3mm}
\label{tauint}
\end{center}
\end{figure}

In contrast to the previous sections, we are now dealing with a
local update algorithm, which is the situation that motivates
this project. As an illustrative example, 
we add a measurement of the integrated autocorrelation 
time $\tau_{\rm int}$ (for the definition, see {\it e.g.}\ 
Ref.\ \cite{Ferenc}) with respect to $Q$. It is expressed in the
number of sweeps (updates of each link variable), 
in a fixed volume with $L=16$. Figure \ref{tauint} shows that 
$\tau_{\rm int}$ increases rapidly as we approach
the continuum limit. 
This confirms that, for the heatbath algorithm,
the problem of ``topological freezing''
becomes severe indeed, once we attain plaquette values of 
$\la P \ra \gtrsim 0.9$.
(The results in Figures \ref{2dU1plat} and \ref{2dU1all} were
obtained by separating the measurements by a number of
sweeps, which clearly exceeds $\tau_{\rm int}$). 

\section{Results for 4d SU(2) Yang-Mills theory}

\newcommand{\gtapprox}{\raisebox{-0.5ex}{$\,\stackrel{>}{\scriptstyle\sim}\,$}}
\newcommand{\ltapprox}{\raisebox{-0.5ex}{$\,\stackrel{<}{\scriptstyle\sim}\,$}}

Finally we study 4d SU(2) Yang-Mills theory, which is
in many respects close to QCD, but computationally cheaper.
Therefore, tests in SU(2) Yang-Mills theory are ideal for hints 
about the practical applicability of the AFHO method to QCD. 
As usual, the lattice gauge field is represented in terms of 
compact link variables $U_{x,\mu} \in {\rm SU(2)}$ \cite{CreutzMM}.

Following Ref.\ \cite{de Forcrand:1997sq}, we determine the 
topological charge density $q_x$, and the topological charge 
$Q = \sum_{x} q_{x}$, by an improved field theoretic definition,
\begin{equation}
\label{EQN001} 
q_{x} [U] =
\frac{1}{16 \pi^2} \sum_{\mu \nu \rho \sigma} \epsilon_{\mu \nu \rho \sigma} 
\sum_{\Box = 1,2,3} \frac{c_\Box}{\Box^4} 
F_{x, \, \mu \nu}^{(\Box \times \Box )}[U] \
F_{x, \, \rho \sigma}^{(\Box \times \Box )}[U] \ ,
\end{equation}
where $F_{x, \, \mu \nu}^{(\Box \times \Box )}$ denotes the dimensionless 
lattice field strength tensor, clover averaged over 
square-shaped Wilson loops of size 
$\Box \times \Box$, and $c_1 = 1.5$, $c_2 = -0.6$, $c_3 = 0.1$.
 
We apply eq.\ (\ref{EQN001}), after performing a number $N_{\rm cool}$
of cooling sweeps with the intention to suppress UV fluctuations in 
the gauge configurations, while preserving the topological structure. 
A cooling sweep amounts to a local minimization of the action, 
{\it i.e.}\ a successive minimization with respect to each gauge link. 
For this minimization we also use an improved version of the 
lattice Yang-Mills action,
\begin{equation}
\label{EQN002} S[U] = \frac{\beta}{16} \sum_x \sum_{\mu \nu} 
\sum_{\Box = 1,2,3}\frac{c_\Box}{\Box^4} \textrm{Tr}
\Big(1 - W_{x, \, \mu \nu}^{(\Box \times \Box )}[U]\Big) \ ,
\end{equation}
where $\beta = 1/(4g^2)$, and $W_{x, \, \mu \nu}^{(\Box \times \Box)}$ 
is a clover averaged loop of size $\Box \times \Box$. Choosing 
an appropriate sweep number $N_\textrm{cool}$ is a 
subtle and somewhat ambiguous task, which will be discussed below.

The lattice action used for the generation of gauge configurations 
is the standard plaquette action, which is obtained from
eq.\ (\ref{EQN002}) by setting $c_1 = 1$, $c_2 = c_3 = 0$.
As in Section 5, the simulations were performed with 
a heatbath algorithm \cite{CreutzMM}, now at $\beta = 2.5$. This 
corresponds to the lattice spacing $a \approx 0.073 \, \textrm{fm}$, 
when the scale is set by identifying the Sommer parameter $r_0$ 
with $0.46\,\textrm{fm}$ \cite{Philipsen:2013ysa}. That value of 
$a$ is in the range of lattice spacings 
$0.05 \, \textrm{fm} \leq a \leq 0.15 \, \textrm{fm}$ typically 
used in contemporary QCD simulations. We generated about 4000 configurations
in each of three volumes, $V = 14^4 \, , \ 16^4 \, , \ 18^4$.
This is also a typical statistics in QCD simulations.

Assigning a topological charge $Q$ to each configuration
leads to the statistics given in Table~\ref{TAB_YM1} for
the sectors $|Q| \leq 4$. We proceed by
computing the correlation function of the 
topological charge density $\langle q_0 q_x \rangle_{|Q|}$ 
in all these sectors $Q$ and volumes $V$. 

The normalization factor on the right-hand-side of
eq.\ (\ref{denseq}) is given by the inverse volume.
As we have observed in the previous sections, the correspondingly
suppressed signal in a large volume is often the bottleneck in the
application of the AFHO method. In order to compensate this 
suppression, which is worrisome in a 4d volume, we now determine
$\la q_{0} q_{x}\ra$ by measuring {\em all-to-all correlations} 
in each configuration, thus taking advantage of the discrete
translational and rotational invariance.

\begin{figure}[h!]
\begin{center}
\includegraphics[angle=270,width=0.55\columnwidth]{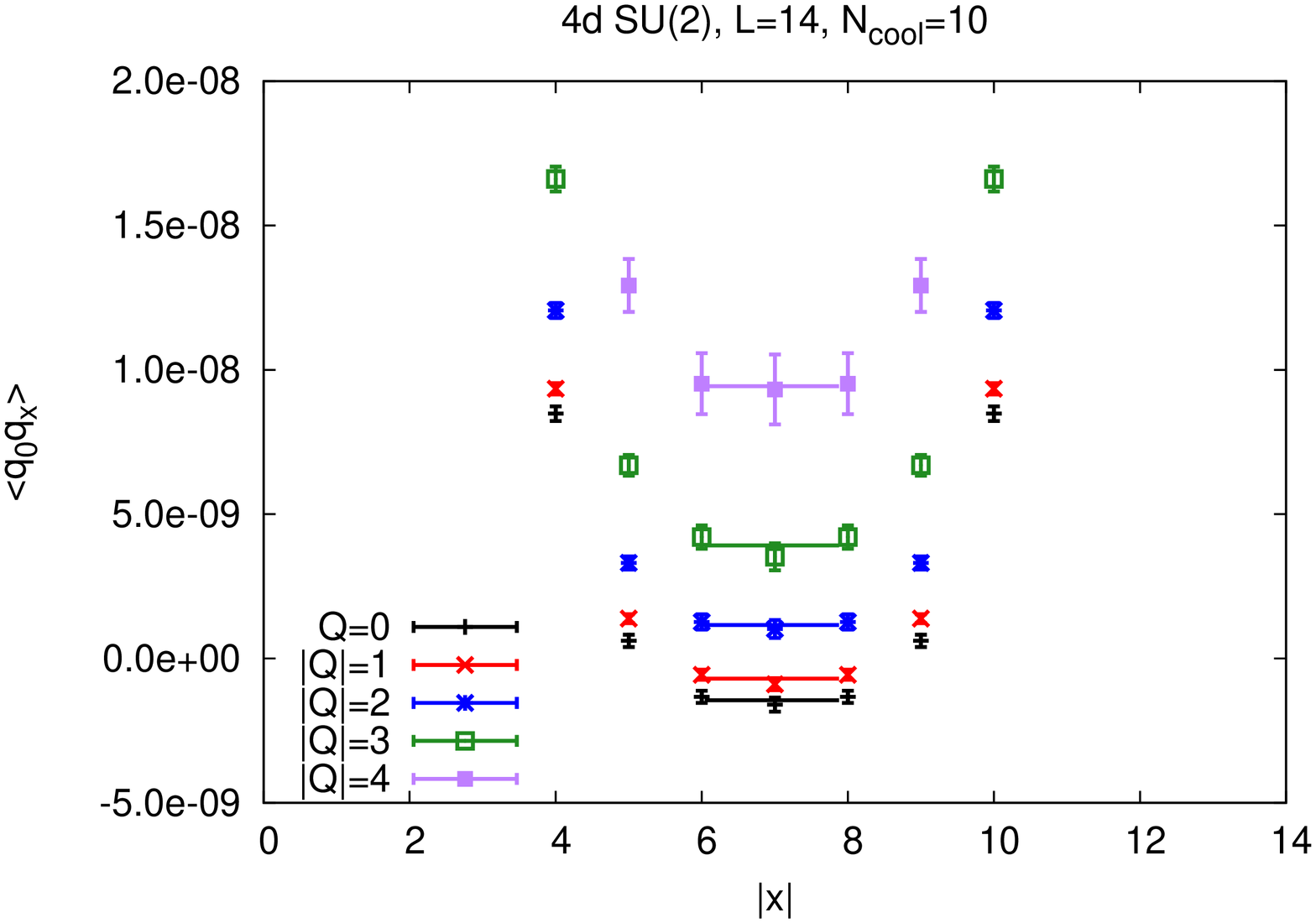}\\
\includegraphics[angle=270,width=0.55\columnwidth]{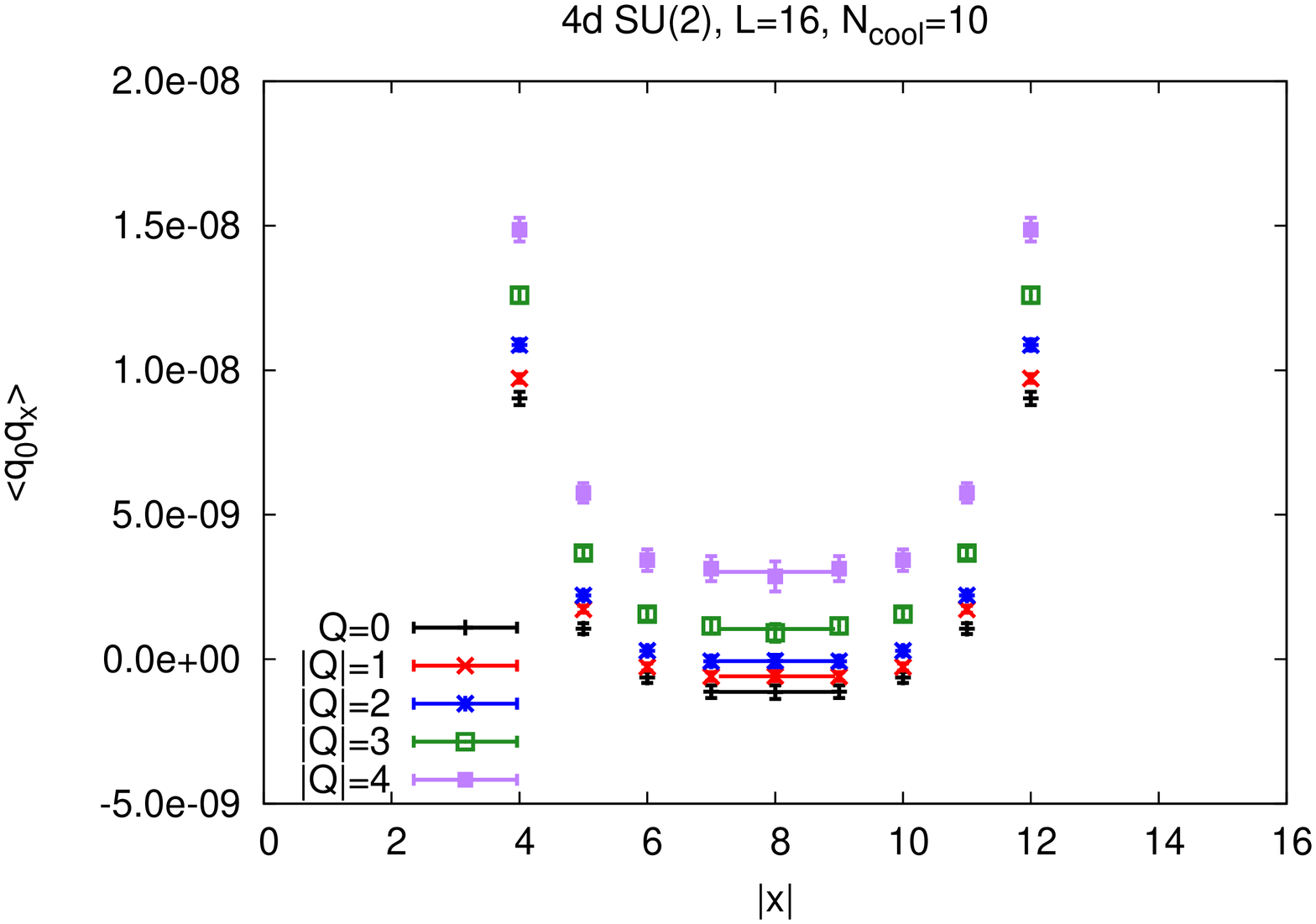}\\
\includegraphics[angle=270,width=0.55\columnwidth]{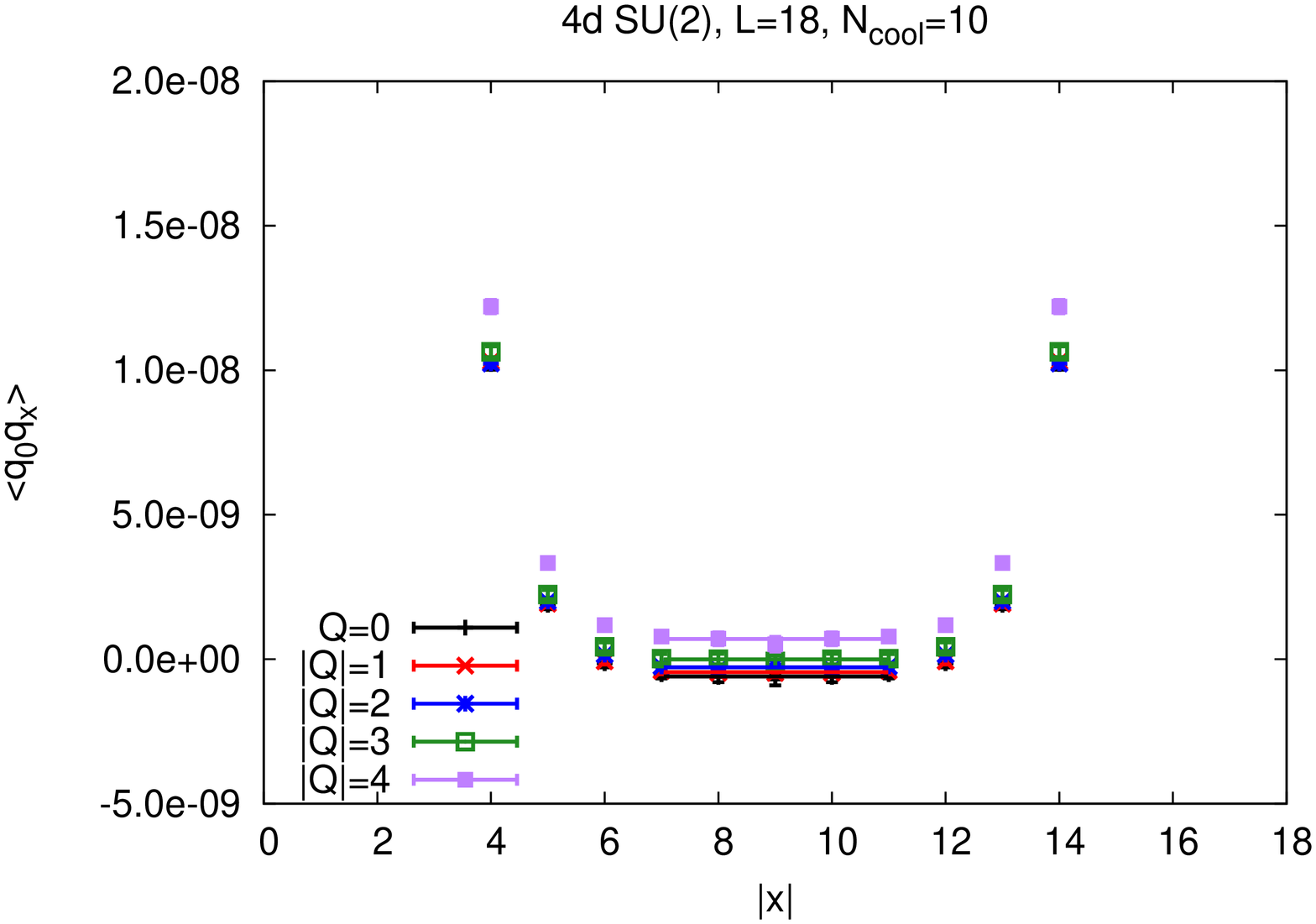}
\caption{\label{FIG_YM1} The correlation $\langle q_0 q_x \rangle_{|Q|}$ 
as a function of the on-axis separation $|x|$, after 
$N_\textrm{cool} = 10$ cooling sweeps, for the 
lattice volumes $V = 14^4 \, , \ 16^4 \, , \ 18^4$. 
Fits of the right-hand-side of eq.\ (\ref{qq1dO2}) with respect 
to $\chi_{\rm t}$ are indicated by the horizontal solid lines.}
\end{center}
\vspace*{-3mm}
\end{figure}

In a second step we fit to these lattice results the right-hand-side 
of eq.\ (\ref{qq1dO2}), {\it i.e.}\
eq.\ (\ref{denseq}) with $c_4 = 0$, with respect to $\chi_{\rm t}$,
at sufficiently large separations $x$, where 
$\langle q_0 q_x \rangle_{|Q|}$ exhibits a plateau.
\begin{table}[h!]
\begin{center}
\begin{tabular}{|c||c|c|c|c|c|}
\hline 
$V$ & $Q=0$ & $|Q|=1$ & $|Q|=2$ & $|Q|=3$ & $|Q|=4$ \tabularnewline
\hline 
\hline 
$14^{4}$ & 1023 & 1591 & \phantom{0}893 & \phantom{0}350 
& \phantom{0}103 \tabularnewline
\hline 
$16^{4}$ & \phantom{0}722 & 1371 & \phantom{0}942 
& \phantom{0}574 & \phantom{0}248 \tabularnewline
\hline 
$18^{4}$ & \phantom{0}622 & 1079 & \phantom{0}898 
& \phantom{0}616 & \phantom{0}402 \tabularnewline
\hline 
\end{tabular}
\caption{\label{TAB_YM1}Number of configurations for three 
volumes $V$, in each topological sector $0 \leq |Q| \leq 4$.
The topological charge has been assigned after 
performing $N_\textrm{cool} = 10$ cooling sweeps.}
\end{center}
\vspace*{-3mm}
\end{table}
Figure~\ref{FIG_YM1} illustrates the determination of $\chi_{\rm t}$ 
after $N_\textrm{cool} = 10$ cooling sweeps, in the 
three lattice volumes under consideration.
Clearly the correlation function $\langle q_0 q_x \rangle_{|Q|}$ 
is different for each topological sector $|Q|$. These differences 
are more pronounced for smaller volumes $V$ and larger topological 
charges $Q$, which shows that the splitting according to eqs.\ 
(\ref{denseq}) and (\ref{qq1dO2}) can indeed be resolved from
our data. Hence the statistics, drastically amplified by the
all-to-all correlations, is indeed sufficient to reveal
the relevant signal.

In particular, for $V = 14^4$ the maximally available on-axis 
separations $|x| = 6, 7, 8$ are at the border-line which
allows us to observe plateaux of $\langle q_0 q_x \rangle_{|Q|}$.
For $V = 16^4$ plateaux are visible in the range $|x| = 7, 8, 9$ and 
in $V = 18^4$ even five points, $7 \leq |x| \leq 11$, 
are consistent with a plateau.

Since the signal, {\it i.e.}\ the differences between the plateau 
values, increases for smaller volumes, a promising strategy might 
be to use anisotropic volumes. For example the volumes of a $14^4$ 
and a $12^3 \times 24$ lattice are similar ({\it i.e.}\ 
both should exhibit a similar signal quality), but the latter 
allows to study larger separations 
(on-axis up to $|x| = 12$, though not with the entire statistics of
all-to-all correlations).

\begin{figure}[h!]
\begin{center}
\includegraphics[angle=270,width=0.49\columnwidth]{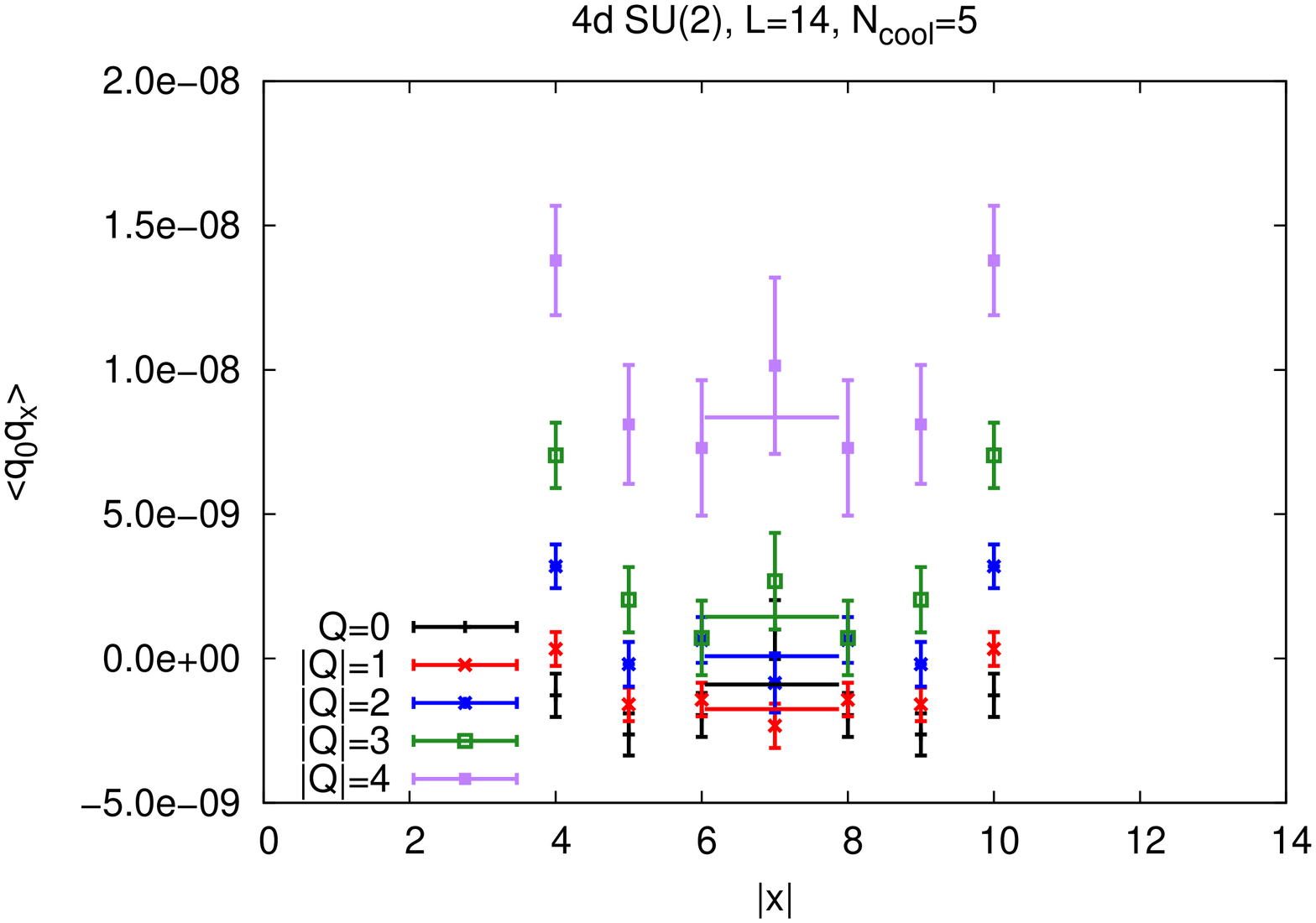}\includegraphics[angle=270,width=0.49\columnwidth]{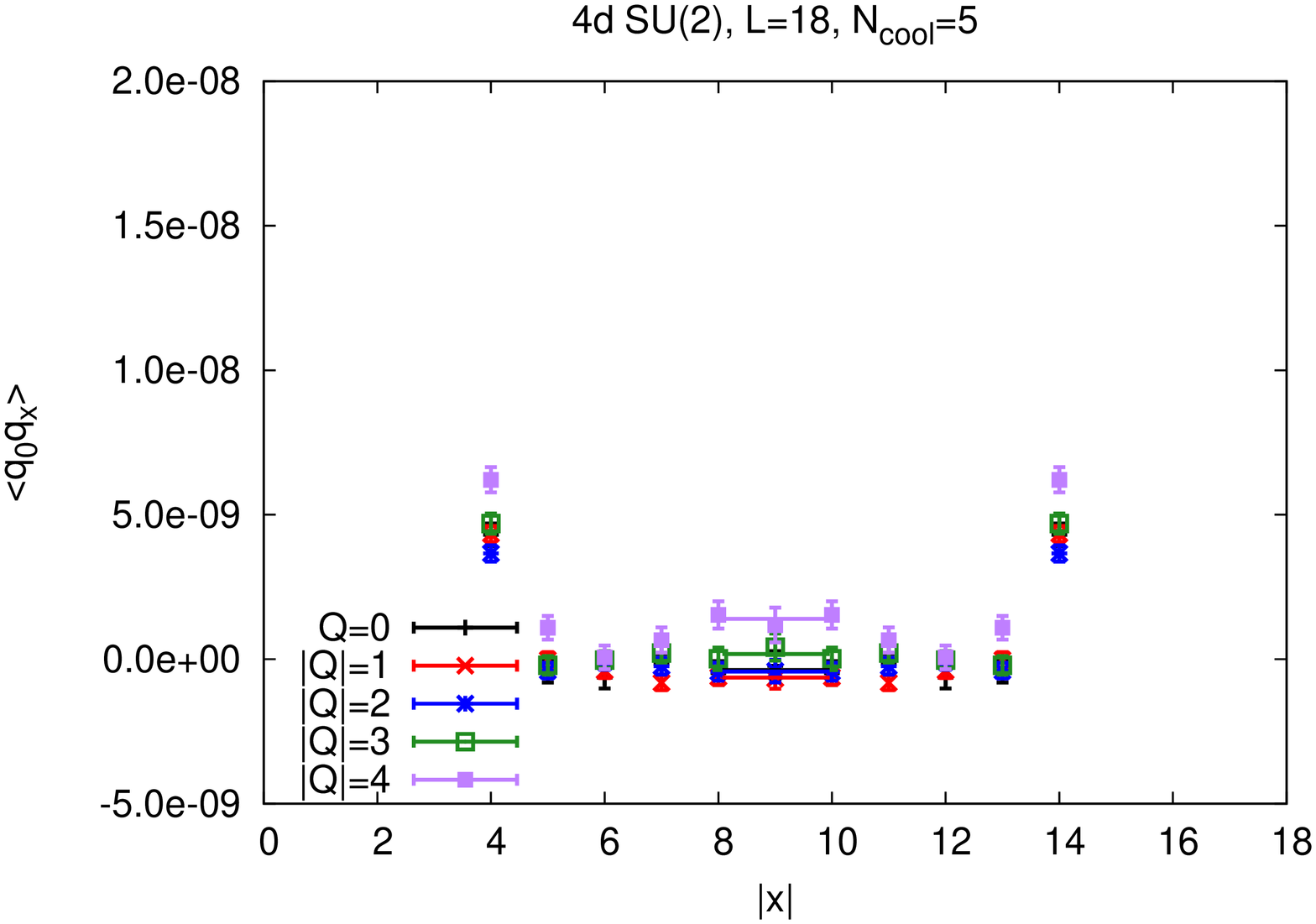}\\
\includegraphics[angle=270,width=0.49\columnwidth]{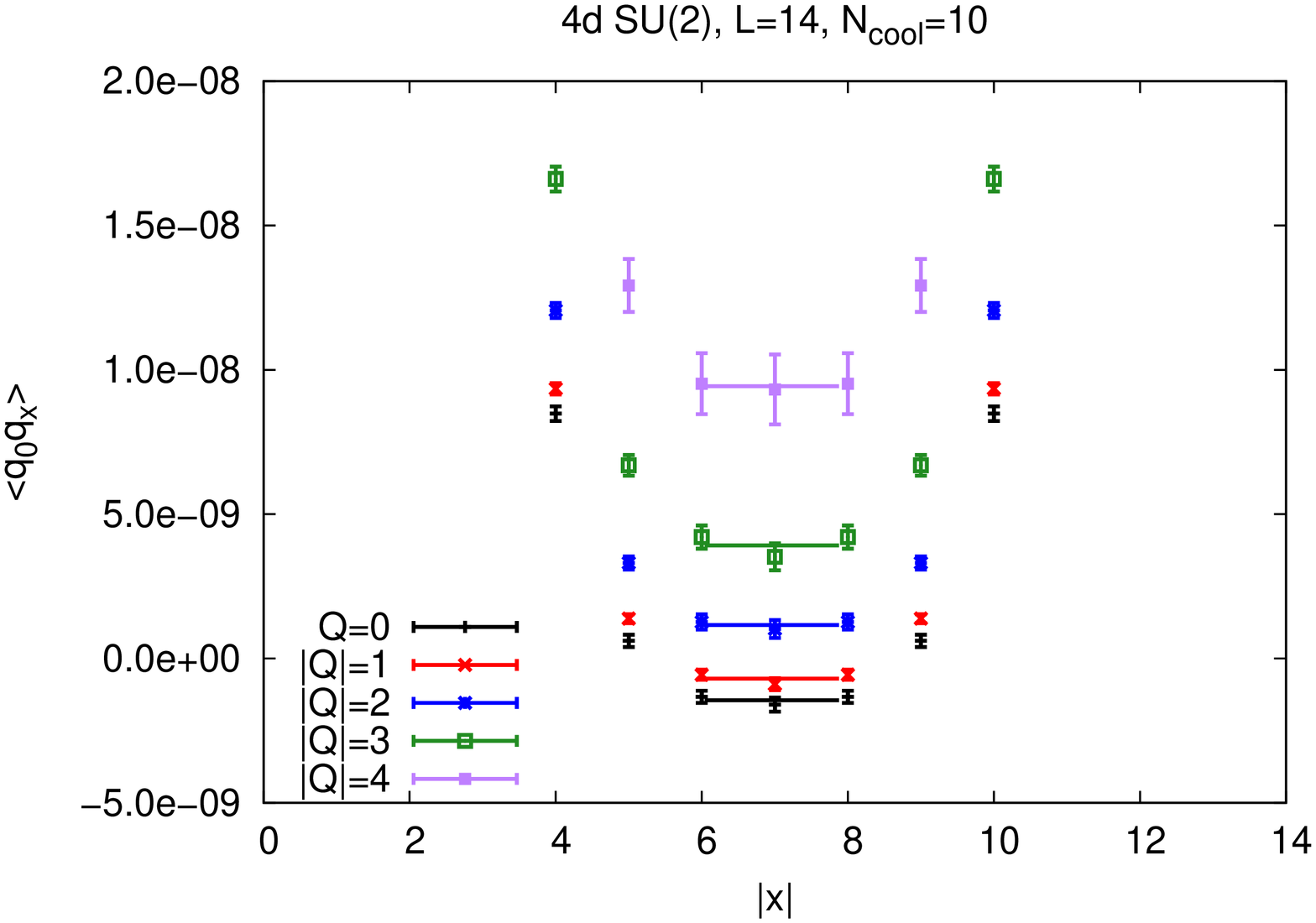}\includegraphics[angle=270,width=0.49\columnwidth]{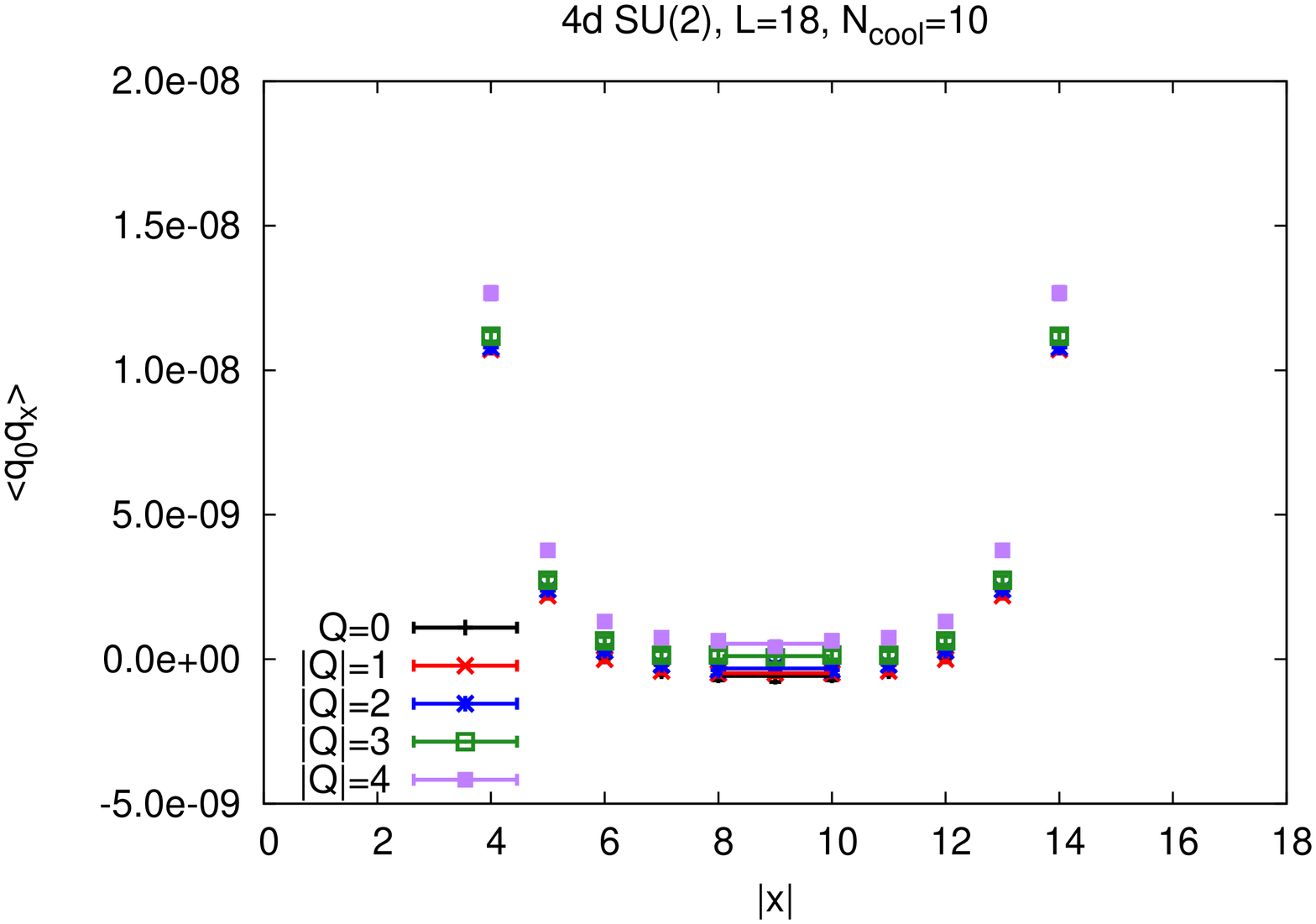}\\
\includegraphics[angle=270,width=0.49\columnwidth]{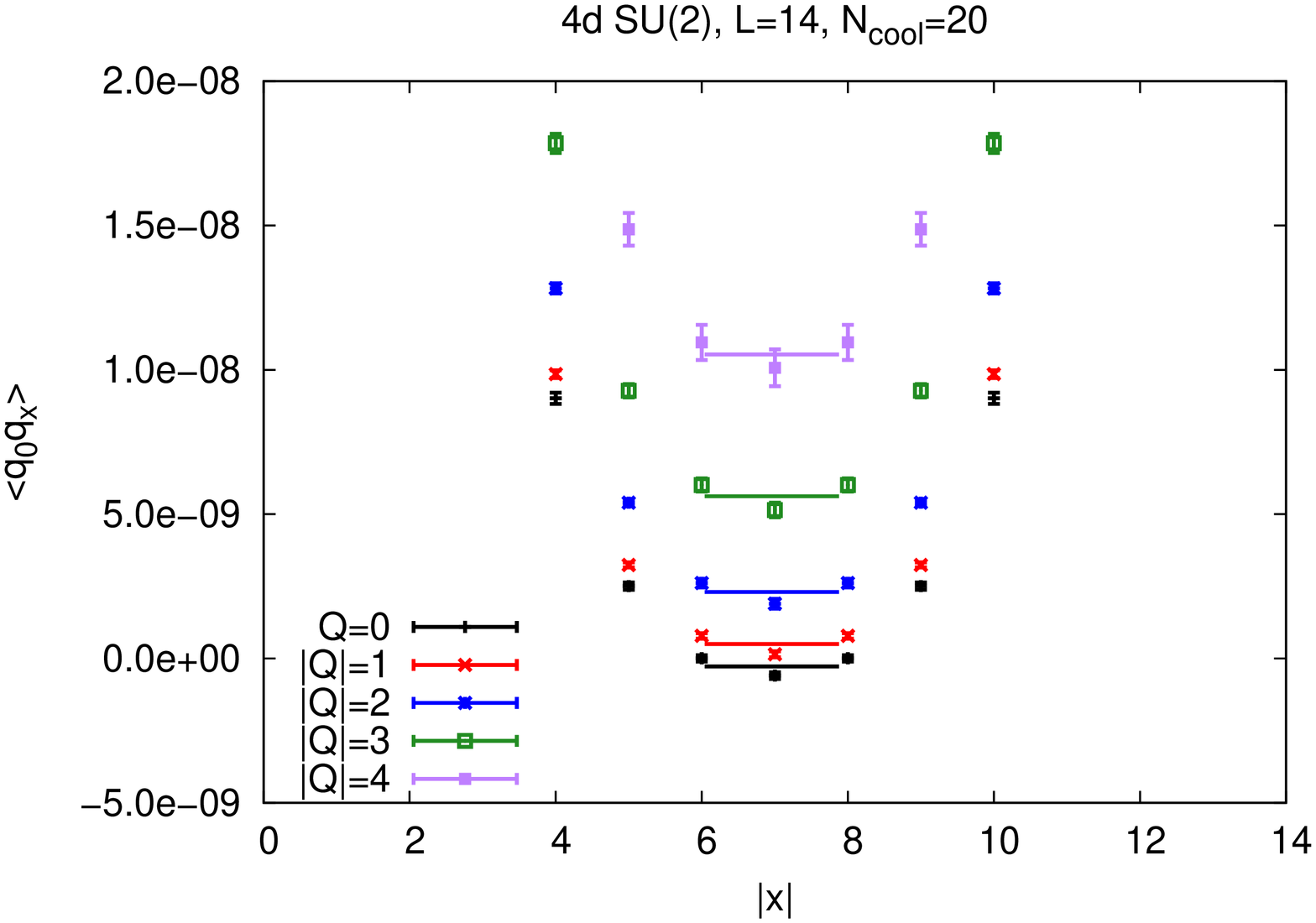}\includegraphics[angle=270,width=0.49\columnwidth]{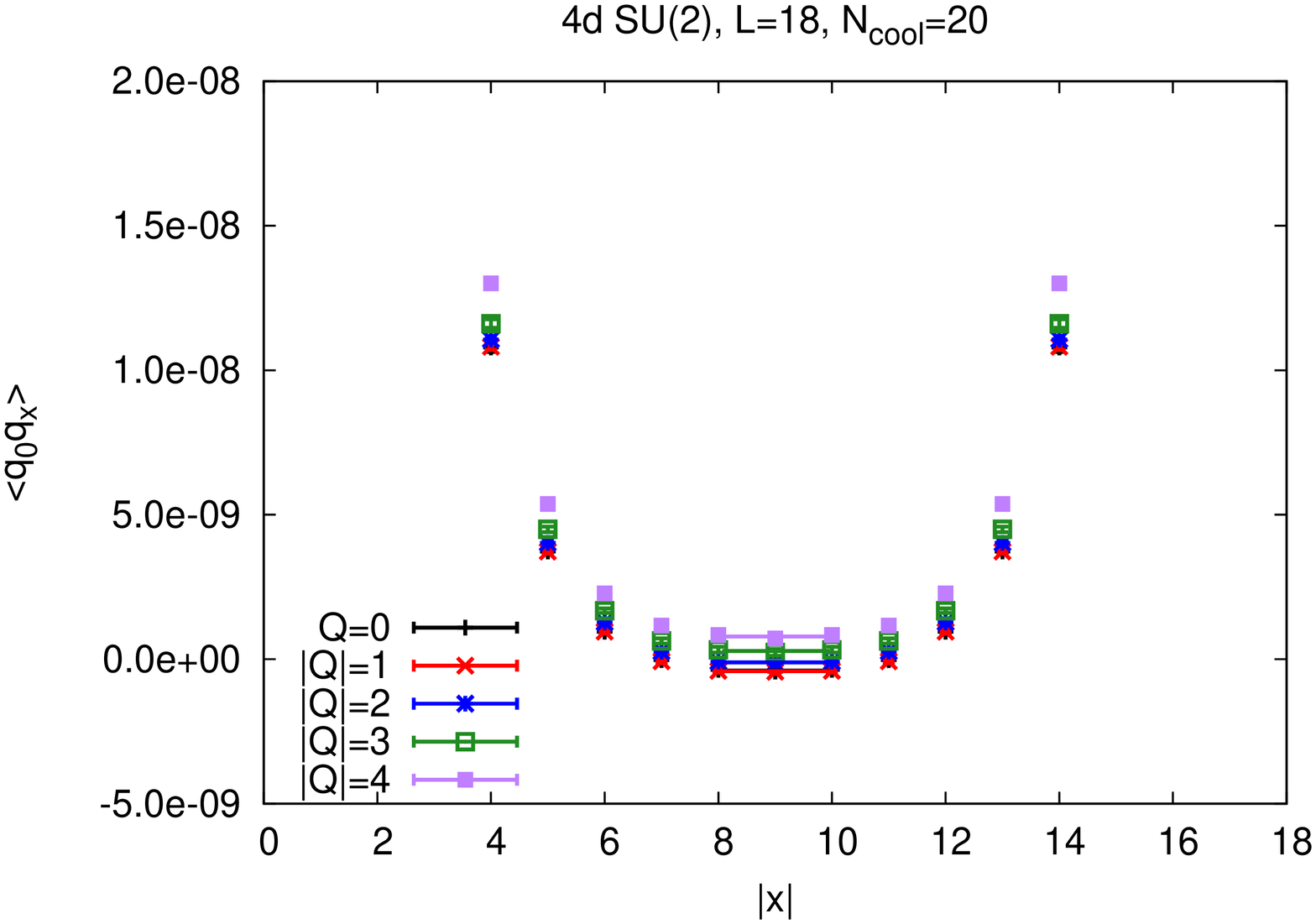}
\caption{\label{FIG_YM2}The correlation $\langle q_0 q_x \rangle_{|Q|}$ 
as a function of the on-axis separation $|x|$ for different numbers 
of cooling sweeps, $N_\textrm{cool} = 5 \, , \ 10 \, , \ 20$, and 
lattice volumes $V = 14^4$ and $18^4$.}
\end{center}
\end{figure}

In Figure~\ref{FIG_YM2} we show determinations of $\chi_{\rm t}$ 
and compare different numbers of cooling sweeps, $N_\textrm{cool} = 
5 \, , \ 10 \, , \ 20$, in the volumes $V = 14^4$ and $18^4$.
For a small number, such as $N_\textrm{cool} = 5$, 
the correlation function $\langle q_0 q_x \rangle_{|Q|}$ is rather 
noisy. This is a consequence of strong UV fluctuations, which are
manifest in the topological charge density $q_x$, and which are 
not filtered out sufficiently at small $N_\textrm{cool}$. For a 
large number of cooling sweeps, like $N_\textrm{cool} = 20$, 
statistical errors are significantly smaller, but the correlation 
function $\langle q_0 q_x \rangle_{|Q|}$ exhibits plateaux only at 
larger separations $|x|$. 

This effect becomes plausible when considering the structure of 
the states contributing to $\langle q_0 q_x \rangle_{|Q|}$. This correlation 
function is the Fourier transform of an analogous correlation 
function, summed over all topological sectors at a finite vacuum angle 
$\theta$ (for a detailed discussion, see Ref.\ \cite{Arthur}). 
The plateau values arise due to the non-vanishing 
vacuum expectation value $ \la q_x \ra $ at $\theta \neq 0$. 
Deviations from these plateaux are predominantly caused by low-lying 
excitations, which correspond to glueballs in Yang-Mills theory. 
Due to the glueball size, the overlap with $q_x | \Omega \rangle$
(where $| \Omega \rangle$ is the vacuum state) increases
when using extensive cooling (then 
$q_x$ is an extended operator resembling a low-lying glueball), 
compared to little or no cooling (then $q_x$ is a highly local operator). 
Consequently, cooling enhances the contribution of excitations 
to the correlation function $\langle q_0 q_x \rangle_{|Q|}$ and, 
hence, causes stronger deviations from the plateaux.

\begin{figure}[h!]
\begin{center}
\includegraphics[angle=270,width=0.55\columnwidth]{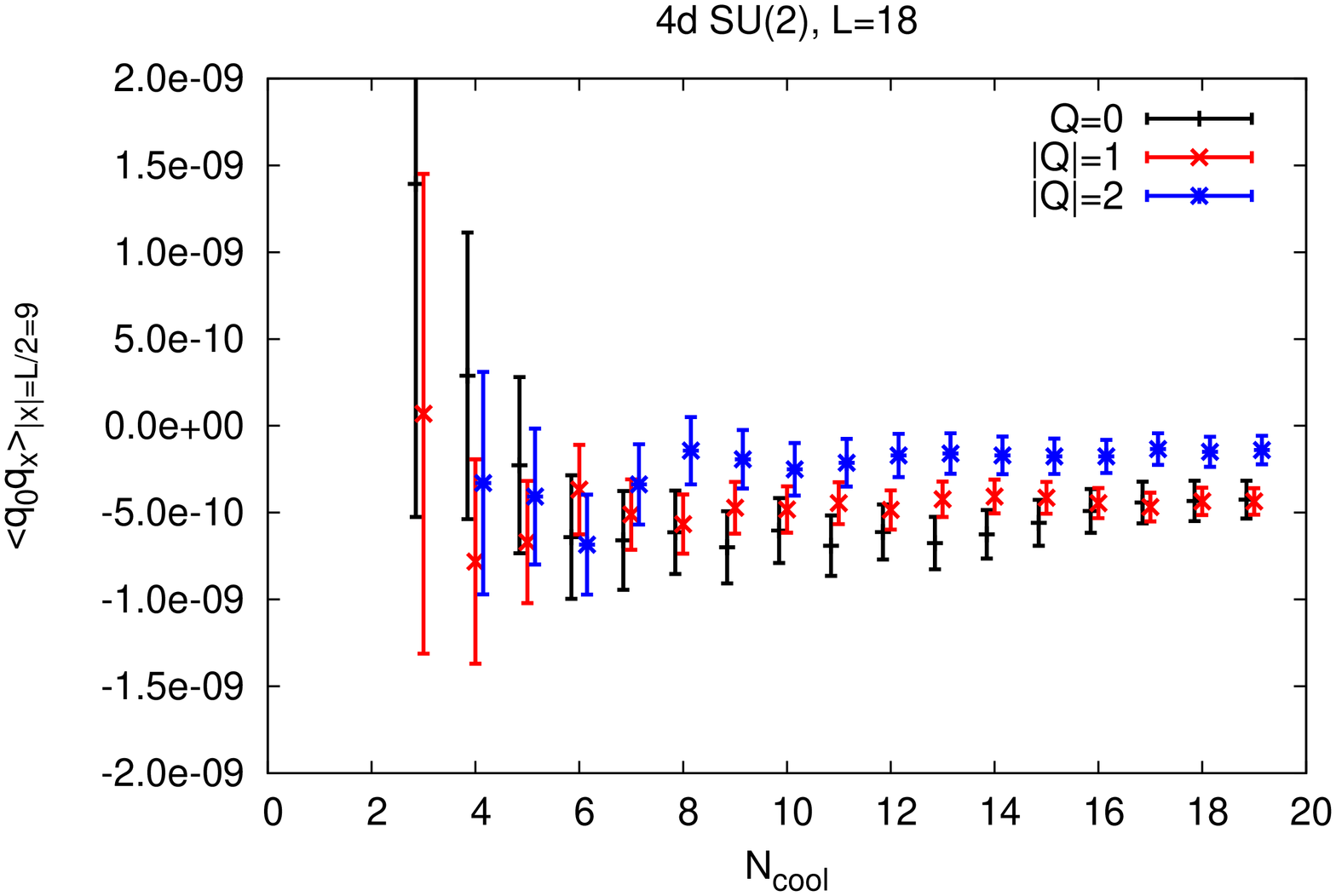}\\
\includegraphics[angle=270,width=0.55\columnwidth]{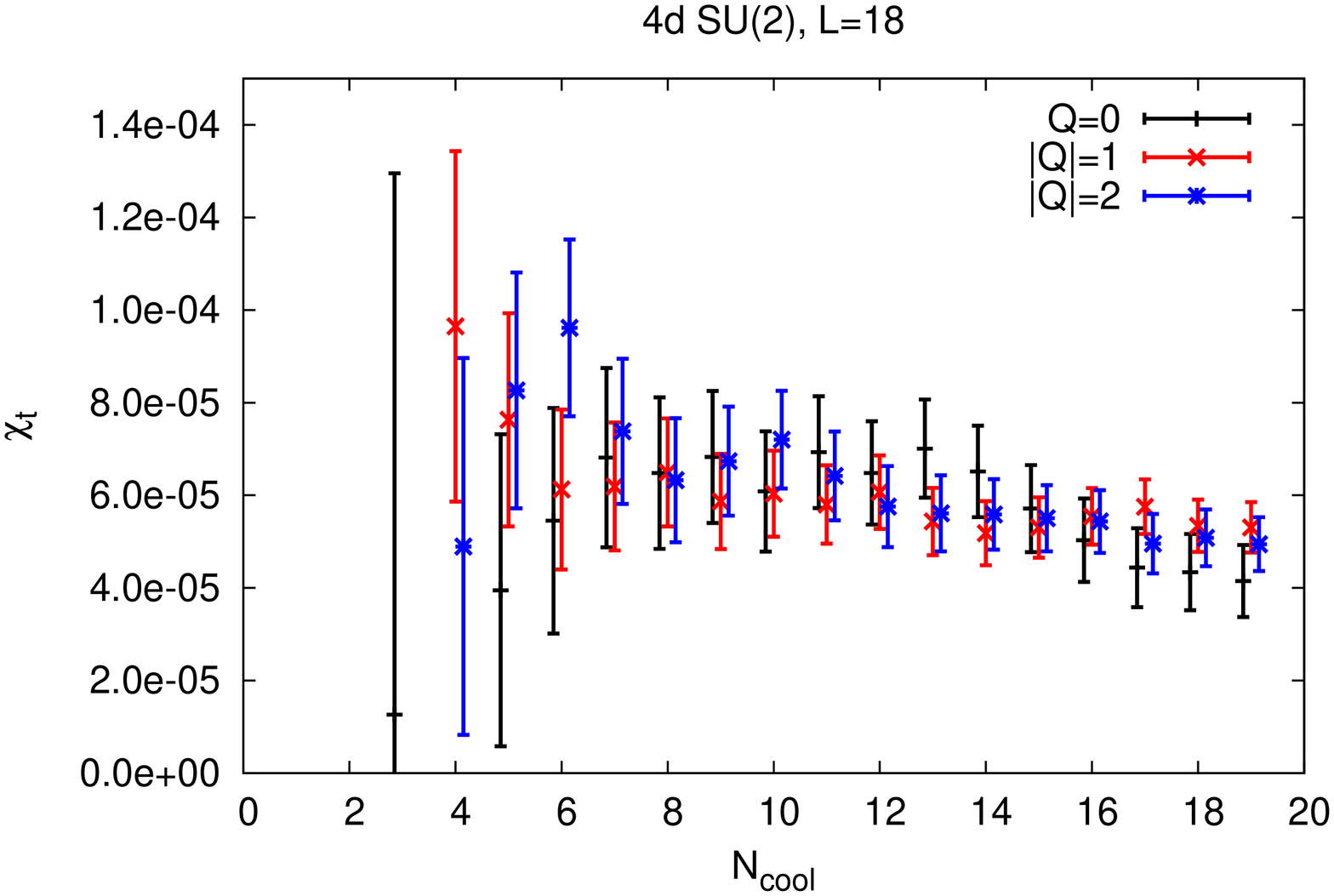}
\caption{\label{FIG_YM3} The correlation $\langle q_0 q_x \rangle_{|Q|}$
at $|x| = L/2 = 9$, and the topological susceptibility $\chi_{\rm t}$,
as a function of the number of cooling sweeps $N_\textrm{cool}$
in the volume $V = 18^4$.}
\end{center}
\end{figure}

In practice one should decide for an optimal compromise, {\it i.e.}\ 
an intermediate number of cooling sweeps. 
Such a compromise can be read off for instance 
from plots showing the dependence of the topological 
susceptibility $\chi_{\rm t}$ (obtained by the AFHO 
method at fixed $|Q|$), or the correlation function 
$\langle q_0 q_x \rangle_{|Q}$ at specific separation $|x| \approx L/2$ 
on $N_\textrm{cool}$. Such plots are shown in Figure~\ref{FIG_YM3} 
for $L = 18$ and fixed topological charge $|Q| = 0, 1, 2$. As 
expected, the statistical errors are quite large for small 
$N_\textrm{cool} \ltapprox 5$. In an intermediate region,
$5 \ltapprox N_\textrm{cool} \ltapprox 15$, there are stable 
plateaux of both $\langle q_0 q_x \rangle_{|Q|}$ and $\chi_{\rm t}$ with 
comparably small statistical errors. The rather long plateaux 
indicate that cooling is a numerically stable procedure, not
destroying topological excitations nor introducing any unwanted 
non-locality effects. At large $N_\textrm{cool} \gtapprox 15$ 
there is a slight trend towards lower $\chi_{\rm t}$, in particular for 
$Q = 0$, which could be a first sign of contamination by excited states. 
An optimal choice for $N_\textrm{cool}$ is somewhere inside the 
plateaux region, such as $N_\textrm{cool} = 8$ or $N_\textrm{cool}$ = 10, 
as we have used in the examples in Figure~\ref{FIG_YM1} and 
Table~\ref{TAB_YM2}.

Numerical results for the larger volumes, 
$V = 16^4$ and $18^4$, and moderate cooling,
$N_\textrm{cool} = 8$ or $10$, where a reasonably 
accurate determination of $\chi_{\rm t}$ seems possible, are 
summarized in Table~\ref{TAB_YM2}. Fits have been restricted 
to sectors $Q$ 
fulfilling $|Q| / (\chi_{\rm t} V) < 0.5$, 
which involves the small expansion parameter in the derivation of 
eq.\ (\ref{denseq}) \cite{AFHO,Arthur}. The resulting 
values for the topological susceptibility agree within the errors with 
a previous straight determination (without topology fixing), which
arrived at $\chi_{\rm t} = 7.0(9) \times 10^{-5}$ \cite{deForcrand:1997sq}.

\begin{table}[h!]
\begin{center}
\begin{tabular}{|c|c||c|c|c|c||c|}
\hline 
$V$ & $N_\textrm{cool}$ & $Q=0$ & $|Q|=1$ & $|Q|=2$ & $|Q|=3$ 
& combined \tabularnewline
\hline 
\hline 
\multirow{2}{*}{$16^{4}$} & \phantom{0}8 & $6.1(9)$ & $5.5(9)$ 
& $7.1(8)$ & & $6.3(6)$ \tabularnewline
\cline{2-6} 
 & 10 & $7.4(10)$ & $5.4(7)$ & $6.5(9)$ & & $6.3(5)$ 
\tabularnewline
\hline 
\hline 
\multirow{2}{*}{$18^{4}$} & \phantom{0}8 & $7.1(14)$ & $~\, 5.8(10)$ 
& $~\, 7.4(12)$ & $9.2(19)$ & $7.3(6)$ \tabularnewline
\cline{2-6} 
 & 10 & $6.2(10)$ & $~\, 5.9(10)$ & $6.6(9)$ & $8.7(11)$ & $7.0(5)$ 
\tabularnewline
\hline 
\end{tabular}
\caption{\label{TAB_YM2}Results for the topological susceptibility 
$\chi_{\rm t} \times 10^5$ extracted from fits to a single sector $|Q|$, 
or combined fits to several sectors ($0 \leq |Q| \leq 2$ for 
$V = 16^4$, $0 \leq |Q| \leq 3$ for $V = 18^4$). A corresponding 
study without topology fixing \cite{deForcrand:1997sq}
obtained $\chi_{\rm t} \times 10^5 = 7.0(9)$.}
\end{center}
\vspace*{-3mm}
\end{table}

We also investigated the magnitude of ordinary finite volume effects, 
(not related to topology fixing). We did this by computing 
and comparing $\chi_{\rm t}$ at unfixed topology using 
$\chi_{\rm t} = \langle Q^2 \rangle / V$. For the three volumes used 
throughout this section, and $N_\textrm{cool} = 10$, we obtain
for $\chi_{\rm t} (V)$
\begin{displaymath}
\chi_{\rm t} (14^{4}) = 7.02(1) \times 10^{-5} , \
\chi_{\rm t} (16^{4}) = 6.97(2) \times 10^{-5} , \
\chi_{\rm t} (18^{4}) = 7.01(3) \times 10^{-5} .
\end{displaymath}
These values agree within the statistical errors.
Moreover, both the differences as well as the statistical errors 
are significantly smaller (by more than a factor of $10$) than the 
uncertainties associated with our $\chi_{\rm t}$ determinations using 
the AFHO method, as listed in Table~\ref{TAB_YM2}. 
Therefore, in our study ordinary finite volume effects 
can safely be neglected.

The similarity of 4d SU(2) Yang-Mills theory and 
QCD suggests that an application of the AFHO method 
to QCD will also allow for a determination of the 
topological susceptibility up to about $10 \, \%$, based on 
$O(1000)$ configurations for each of the topological 
sectors with small $|Q|$. We plan to explore this scenario
in the near future.

\section{Conclusions}

We have investigated the AFHO method \cite{AFHO}
for the evaluation of the topological susceptibility $\chi_{\rm t}$
based on the correlation of the topological charge density.
Amazingly, this method allows --- in principle --- for a measurement
of $\chi_{\rm t}$ even within a fixed topological sector, and
low $|Q|$ are most promising.

We have seen that the method as such works; 
in some cases it provided results, which are correct to 2 or 3 
digits. Hence situations do exist, where the approximations in the
derivation of formula (\ref{qq1dO2}) (in particular including
order $O(1/V^{2})$ incompletely, and neglecting all 
terms of $O(1/V^{3})$) can be justified, as we
could confirm on the non-perturbative level.

This approximation neglects the kurtosis term in relation
(\ref{denseq}), and all higher terms. They are 
suppressed by the inverse volume. Moreover, it is a generic
property that topological charges are approximately Gauss
distributed, such that $|c_{4}|$ is small. As a peculiar case,
it vanishes in the continuum 1d O(2) model, and in that case 
there are no higher corrections either (cf.\ Appendix A).

Regarding the limitations of the applicability range, we note
that --- for most models studied here --- the theoretical condition 
of measuring the correlation ``at large separation''
turned out not to be worrisome in practice; only in Section 6
this issue has some relevance. However, the
AFHO method generally runs into trouble when the volume $V$ 
increases. Then the signal in $\la q_{0} \, q_{x} \ra_{|Q|}$ 
is suppressed by a factor $1/V$, and the separation between
the predicted plateaux by $1/V^{2}$.
If we want the correlation length to be clearly larger
than the lattice spacing (so that lattice artifacts are
under control), we need a sizable lattice to keep
the finite size effects under control as well. With these
conflicting requirements, even in our 2d test models, and despite 
a statistics of $10^{7}$ measurements, this method led to results
with rather large statistical errors. 

Thus we observe that
the AFHO method, which is based on topologically restricted
measurements, is plagued by unusually persistent finite size
effects. In usual settings, the latter are exponentially
suppressed, {\it i.e.}\ they are 
$\propto \exp (-{\rm const.} \, L / \xi)$, and quite small
if $L / \xi \gtrsim 4$. However, topologically
restricted numerical measurements are very sensitive to
finite size effects: we recall that relation (\ref{denseq})
is a truncated polynomial expansion in $1/V$, but we can
also refer directly to the large ratios $L/\xi$, which were
used all over this study: {\it e.g.}\ in the upper plot of
Figure \ref{1dO2B4qq} it amounts to $L/\xi > 22$, which
usually makes finite size effects negligible, but here they are
significant. This property calls for a larger size $L$. 
In turn, that causes problems in extracting the subtle effect,
which is relevant for the indirect determination of $\chi_{\rm t}$.

That issue might be the bottleneck for the prospects to apply the 
AFHO method in four dimensions, and it has inspired the approach
of restriction to sub-volumes \cite {JLQCD14}.
However, our study in Section 6 shows that the suppression of the 
wanted signal by the inverse volume can be successfully compensated
if we enhance the statistics by means of all-to-all correlation 
measurements.
We have seen in 4d SU(2) Yang-Mills theory that this procedure 
does provide sufficient precision for $\la q_{0} q_{x}\ra$, even for
a moderate number of $O(1000)$ configurations in one topological
sector. Due to this observation, the AFHO method in its original
form appears quite promising in QCD. We are going to test if it
enables also there the evaluation of $\chi_{\rm t}$ in an 
unconventional way, to an accuracy of about $10 \, \%$.

\ \\
\noindent
{\bf Acknowledgements} \ Lilian Prado has worked on this project
at an early stage. We also thank Christopher Czaban
and Philippe de Forcrand for interesting
discussions. This work was supported in part by the 
Mexican {\it Consejo Nacional de 
Ciencia y Tecnolog\'{\i}a} (CONACYT) through project 155905/10,
by DGAPA-UNAM, grant IN107915,
and by the {\it Helmholtz International Center for FAIR} within
the framework of the LOEWE program launched by the State of Hesse.
A.D.\ and M.W.\ acknowledge support by the Emmy Noether Programme of the
DFG (German Research Foundation), grant WA 3000/1-1, and C.P.H.\ 
acknowledges support through the project {\it Redes Tem\'{a}ticas de 
Colaboraci\'{o}n Acad\'{e}mica 2013,} UCOL-CA-56.
Calculations on the LOEWE-CSC and on the FUCHS-CSC high-performance 
computer of the Frankfurt University were conducted for this project. 
We would like to thank HPC-Hessen, funded by the State Ministry 
of Higher Education, Research and the Arts, for
programming advice.

\appendix

\section{Topological charge density correlation in the
continuum 1d O(2) model}

The key quantity of this work is the correlation function of 
the topological charge density. In this appendix we compute it 
analytically for the 1d O(2) model, formulated in continuous
Euclidean time $x$, with periodicity length $L$.
For this purpose, it is useful to include a $\theta$-term
in the action,
\be
S [\vp ] = \frac{\beta_{\rm cont}}{2} \int_{0}^{L_{\rm cont}} 
dx \, \vp '(x)^{2} - \ri \theta Q [\vp ] \ .
\ee
In the canonical formulation of quantum mechanics,
the corresponding Hamilton operator, its energy eigenfunctions
and eigenvalues read \cite{rot97}
\be
\hat H = \frac{1}{2 \beta_{\rm cont}} \ \Big( \hat p - 
\frac{\theta}{2 \pi} \Big)^{2} \ , \
\la \vp | n\ra = \frac{1}{\sqrt{2 \pi}} \ e^{\ri n \vp} \ , \
E_{n} = \frac{1}{2 \beta_{\rm cont}} 
\Big( n - \frac{\theta}{2 \pi} \Big)^{2} \ , \
\ee
where $\hat p = - \ri \frac{\partial}{\partial \vp}$ and
$n \in \Z$.
The operator for the topological charge density is given by
\bea
\hat q &=& \frac{1}{2 \pi} [\hat H, \hat \vp ] =
\frac{1}{2 \pi \beta_{\rm cont}} \Big( -\frac{\partial}{\partial \vp}
+ \ri \frac{\theta}{2\pi} \Big) \ . 
\eea
This operator is {\em anti-Hermitian} (due to the Euclidean
time derivative of the Hermitian operator $\hat \vp$),
with the matrix elements
\be
\la m | \hat q | n \ra = \frac{1}{(2 \pi)^{2} \beta_{\rm cont}}
\int_{-\pi}^{\pi} d \vp \, e^{- \ri m \vp} \,
\Big( -\frac{\partial}{\partial \vp} + \ri \frac{\theta}{2\pi} \Big)
\, e^{\ri n \vp} = \frac{\ri ( \theta - 2 \pi n)}{(2 \pi)^{2} \beta_{\rm cont}} 
\, \delta_{mn} \ . \nn
\ee
Hence the expectation value
\be
\la q(x) \ra = \frac{1}{Z(\theta )} \frac{\ri}{(2 \pi)^{2} \beta_{\rm cont}} 
\sum_{n \in \Z} ( \theta - 2 \pi n) \, e^{-E_{n}L_{\rm cont}} \ ,
\ {\rm with} \ Z(\theta ) = \sum_{n \in \Z} \, e^{-E_{n}L_{\rm cont}}
\nn
\ee
is imaginary in general (of course it vanishes at $\theta =0$).

Therefore the corresponding correlation function is in general
{\em negative,}\footnote{In field theoretic models, the correlation 
of the topological charge density over large distance is 
known to be negative as well \cite{Seiler}.}
\bea
\la q(0) q(x) \ra &=& \frac{1}{Z (\theta )}
\sum_{m,n \in \Z} \la m | \hat q | n \ra \, \la n | \hat q | m \ra
\, e^{-E_{n}x - E_{m} \, (L_{\rm cont} - x)} \nn \\
&=& - \frac{1}{Z (\theta )} \frac{1}{(2 \pi)^{4} \beta^{2}_{\rm cont}}
\, \sum_{n \in \Z} ( \theta - 2 \pi n )^{2} \, e^{-E_{n}L_{\rm cont}} \ .
\eea
It is remarkable that this correlation is independent of $x$,
if $x/L_{\rm cont} \notin \Z$ (this condition allows us to insert a unit
factor $\sum_{n} | n \ra \la n |$ between the end-points).

The vacuum angle $\theta$ enables also the computation of the
topologically restricted partition function,
\be
Z_{Q} = \frac{1}{2 \pi} \int_{-\pi}^{\pi} d \theta \,
Z(\theta ) \, e^{-\ri Q \theta} = \frac{1}{2 \sqrt{\pi \alpha}} 
e^{-Q^{2}/(4 \alpha) } \ ,
\ \ \alpha = \frac{L}{8 \pi^{2} \beta_{\rm cont}} \ , \nn
\ee
and correlation function,
\bea
\la q(0) q(x) \ra_{Q} &=& \frac{1}{2 \pi Z_{Q}} \int_{-\pi}^{\pi} d \theta \,
Z (\theta ) \la q(0) q(x) \ra \, e^{-\ri Q \theta} \nn \\
&=& - \frac{1}{(2 \pi)^{5} \beta^{2}_{\rm cont} Z_{Q}}
\int_{-\infty}^{\infty} d \theta \, \theta^{2} \,
e^{-\alpha \theta^{2} - \ri \theta Q} \nn \\ 
&=& \frac{1}{32 \pi^{4} \beta^{2}_{\rm cont} \alpha}
\Big( -1 + \frac{Q^{2}}{2 \alpha} \Big) \ .
\eea
Finally we insert $\alpha$ and $\chi_{\rm t} =  \alpha L_{\rm cont}/2$ 
(cf.\ Table \ref{xichit})
to arrive at
\be
\la q(0) q(x) \ra_{Q} = - \frac{\chi_{\rm t}}{L_{\rm cont}}
+ \frac{Q^{2}}{L^{2}_{\rm cont}} \ .
\ee
Also the topologically restricted correlation function
is constant in $x$, which explains that the data in Section 3 
attain the plateau values immediately.

Moreover, we see that eq.\ (\ref{qq1dO2}) is {\em exact} in this
specific case, which is consistent with the fact that the kurtosis
vanishes \cite{rot97}. Therefore, in our numerical study presented in
Section 3, the actual issues are lattice artifacts and the visibility
of the predicted plateau values in numerical simulation data; both
are generally relevant questions.

\end{document}